%% file: main.tex
\documentclass[aps,prl,reprint,superscriptaddress,floatfix]{revtex4-2}
\ifdefined\pdfminorversion\pdfminorversion=7\relax\fi
\usepackage{amsmath,amssymb,bm,graphicx}
\usepackage[colorlinks=true,linkcolor=black,citecolor=blue,urlcolor=blue]{hyperref}

\newcommand{\sgn}{\operatorname{sgn}}

\begin{document}
\title{Revealing the Hidden Symmetry of the Asymmetric Quantum Rabi Model}
\author{Zi-Min Li}
\author{Zi-Cheng Liu}
\author{Long-Jie Li}
\author{Xuan Zou}
\author{Jin-Xiang Fu}
\affiliation{Institute of Quantum Physics, Hunan Key Laboratory of Nanophotonics and Devices, Hunan Key Laboratory of Super-Microstructure and Ultrafast Process, School of Physics, Central South University, Changsha 410083, China}

\author{Xilin Lu}
\author{Murray T. Batchelor}
\email{murray.batchelor@anu.edu.au}
\affiliation{Mathematical Sciences Institute, The Australian National University, Canberra, ACT 2601, Australia}
\date{\today}

\begin{abstract}
A longitudinal bias breaks ordinary parity in the asymmetric quantum Rabi model, yet exact crossings reappear when the bias equals an integer multiple $m$ of the oscillator energy spacing.
We trace the hidden symmetry to exchange between equal-energy partners $m$ rungs apart on two displaced oscillator ladders.
Tunneling to other rungs makes this exchange vary in time.
We construct the exact symmetry operator by adding spin and oscillator coherences that cancel this variation.
At nonzero coupling and tunneling, the resulting charge has no zero modes, and its sign divides the full Hilbert space into two generalized-parity sectors that cannot mix.
Spin-resolved mirror interference probes coherence between the two opposite-parity states at a crossing.
Its small-displacement exponent gives $m$, and its leading coefficient, together with the charge magnitude, predicts the mixing induced by a slight change in bias.
\end{abstract}
\maketitle

When a symmetry prevents two states from mixing, their energy levels can cross as a parameter is varied.
Breaking that symmetry usually turns the crossing into an avoided one.
In the quantum Rabi model (QRM), which describes a two-level system coupled to a single oscillator~\cite{Rabi1936,Rabi1937,JaynesCummings1963}, ordinary parity separates states that are even or odd under a combined oscillator reflection and spin transformation~\cite{Braak2011,ChenBOA2012,Zhong2013,XieReview2017}.
A longitudinal bias breaks this parity~\cite{LuPTQRM2023}, and levels that previously crossed generally repel.
Yet exact crossings reappear at selected couplings when the bias is an integer multiple of the oscillator energy spacing~\cite{Zhong2014,LiBatchelor2015,LiBatchelor2016,Wakayama2017,SempleKollar2018,Ashhab2020,Kimoto2021}.
Similar crossings occur in other biased models~\cite{LuDicke2021,XieChenTwoPhoton2021,XieChenTwoPhoton2022,YanTwoMode2022,ReyesBustosTwoPhoton2026}, including anisotropic and Rabi--Stark variants~\cite{LiBatchelor2021,Lu2022}.
What hidden symmetry replaces ordinary parity in the integer-biased QRM, and what does it conserve during evolution~\cite{LowKochAshhab2026}?

Previous work has approached this puzzle from two directions: the symmetry operator and the oscillator states involved in the crossings.
Algebraic constructions give a commuting generator at integer bias and its normalization to a parity operator~\cite{Mangazeev2021,ReyesBustos2021,XieChen2022}.
Displaced oscillators~\cite{Irish2007,LiGAA2021} account for selective tunneling~\cite{LiBatchelor2021,ChenSimulation2024,Song2026}, while level matching in a double well explains crossings at large qubit splitting~\cite{YangChenLuo2025}.
These two views meet at zero tunneling: a recent closed-form generator~\cite{KollarDutta2026} exchanges equal-energy partners on the displaced oscillator ladders.
What remains unclear is how to build a conserved charge from this pairwise exchange once each partner can tunnel to many other levels.
We construct the charge by canceling the bare exchange's variation with spin and oscillator coherences.
Its sign defines two sectors whose probabilities stay fixed as spin populations evolve.
At an isolated crossing, spin-resolved interference with a shifted mirror image probes the same exchange; its amplitude and the charge magnitude predict mixing under weak bias detuning.

\begin{figure}[!tbp]
\centering
\includegraphics[width=\columnwidth]{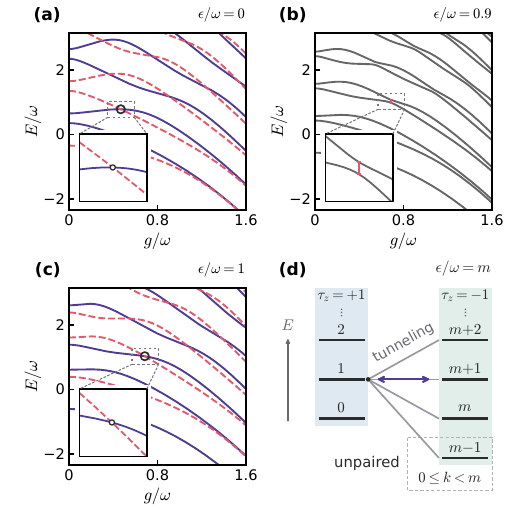}
\caption{Disappearance and revival of level crossings.
(a--c) Spectra at $\Delta/\omega=0.7$, $\epsilon/\omega=0,0.9,1$.
Purple solid and coral dashed curves denote positive and negative parity, ordinary in (a) and hidden in (c); dark gray levels in (b) have no parity label.
Circles mark first-family crossings; the coral bar gives the minimum gap ($\simeq0.09\omega$).
Insets (b,c) share the same coupling and energy window.
(d) Gray lines show allowed tunneling from $n=1$, including to unpaired levels (box); only $k=m+1$ is its equal-energy partner (double arrow).}
\label{fig:revival}
\end{figure}

We use the convention of Ref.~\cite{LiBatchelor2021},
\begin{equation}
 H_{\rm A}=\omega a^\dagger a+g\sigma_x(a+a^\dagger)
       +\frac{\epsilon}{2}\sigma_x+\frac{\Delta}{2}\sigma_z,
 \label{eq:model}
\end{equation}
with $\hbar=1$, oscillator frequency $\omega>0$, coupling $g$, bias $\epsilon$, and qubit splitting $\Delta$.
Write $h=H_{\rm A}/\omega$, $\alpha=g/\omega>0$, and $\delta=\Delta/(2\omega)$.
The spin axes along and across the displaced branches are $\tau_z=\sigma_x$ and $\tau_x=\sigma_z$, with $\tau_y=-\sigma_y$.
The splitting term flips these branches and therefore acts as tunneling.
Oscillator reflection is $P_a=(-1)^{a^\dagger a}$; ordinary QRM parity is $P=\sigma_zP_a$.
Figure~\ref{fig:revival}(a)--(c) shows its breaking and the reappearance of crossings.
We consider $\epsilon=m\omega$, $m=0,1,\ldots$, initially with $\delta>0$; negative splitting follows by a fixed spin transformation.

At zero tunneling, the two $\tau_z$ branches are independent oscillators displaced by $\mp\alpha$.
Their states are $\mathcal D(\mp\alpha)|n\rangle|\pm\rangle$, where $\mathcal D(s)=\exp[s(a^\dagger-a)]$, and their energies are $E_n^\pm/\omega=n-\alpha^2\pm m/2$.
Upper level $n$ and lower level $n+m$ have the same energy: their amplitudes accumulate the same dynamical phase, leaving their relative phase fixed.
The lowest $m$ levels on the lower ladder have no such partners.
When tunneling is turned on, each partner can reach many levels on the opposite ladder [Fig.~\ref{fig:revival}(d)].

To compare these amplitudes in one oscillator basis, we reflect the oscillator only on the $\tau_z=-1$ branch.
This conditional reflection, $T=\operatorname{diag}(1,P_a)$, brings both centers to $-\alpha$.
Writing $b=a+\alpha$ and $|n\rangle_b=\mathcal D(-\alpha)|n\rangle$, we obtain
\begin{equation}
 \widetilde h=ThT^\dagger
 =b^\dagger b-\alpha^2+\frac m2\tau_z+\delta P_a\tau_x .
 \label{eq:center}
\end{equation}
The states $|n,\pm\rangle_b=|n\rangle_b|\pm\rangle$ form the two columns of a displaced Fock lattice~\cite{CaiWang2021}.
Tunneling now flips the spin and reflects the oscillator: the amplitude from $|n,+\rangle_b$ to $|k,-\rangle_b$ is $\delta\,{}_b\langle k|P_a|n\rangle_b$.
The same reflection fixes the relative signs of all these amplitudes.

To exchange the equal-energy partners, raise the spin and remove $m$ oscillator quanta, or reverse both steps:
\begin{equation}
 C_{m,0}=b^m\tau_++(b^\dagger)^m\tau_- ,
 \label{eq:bare_exchange}
\end{equation}
where $\tau_\pm=(\tau_x\pm i\tau_y)/2$.
The spin and oscillator energy changes cancel, making this bare exchange conserved at zero tunneling.
Its expectation sums the real coherences between partners, weighted by their oscillator matrix elements.
Within each pair, the sum and difference of the partners have opposite exchange signs:
\begin{equation}
 \begin{aligned}
 &|\phi_{n,\pm}\rangle
 =\frac{|n,+\rangle_b\pm|n+m,-\rangle_b}{\sqrt2},\qquad n\ge0,\\
 &C_{m,0}|\phi_{n,\pm}\rangle
 =\pm\sqrt{\frac{(n+m)!}{n!}}\,|\phi_{n,\pm}\rangle .
 \end{aligned}
 \label{eq:pair_basis}
\end{equation}
The oscillator matrix element fixes the magnitude; the relative phase fixes the sign.
At $m=0$, undoing $T$ gives ordinary parity $P$.

\begin{figure}[!tbp]
\centering
\includegraphics[width=\columnwidth]{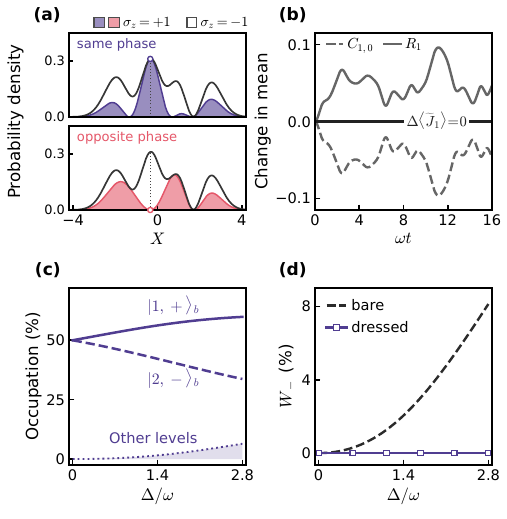}
\caption{Exchange, compensation, and dressing at $m=1$, $g/\omega=0.7$; $X=(a+a^\dagger)/\sqrt2$.
(a) Zero-tunneling densities of $T^\dagger|\phi_{1,+}\rangle$ (top) and $T^\dagger|\phi_{1,-}\rangle$ (bottom).
The black outline gives the total density, with $\sigma_z=+1$ filled and $\sigma_z=-1$ white beneath it.
Circles mark the $\sigma_z=+1$ density in each state where the partner amplitudes agree ($X\simeq-0.343$).
(b) At $\Delta/\omega=0.7$ and initial relative phase $\pi/3$, changes in bare exchange $C_{1,0}$ and reflection correction $R_1=\delta P_a/(2\alpha)$ cancel: $\Delta\langle\widetilde J_1\rangle=0$.
(c) Dressed same-phase state: occupations of the bare partners $|1,+\rangle_b$ and $|2,-\rangle_b$ and the combined occupation of all other ladder levels.
(d) Negative-parity weight $W_-$ of the fixed bare same-phase state and its dressed continuation, using $\mathcal P_-$ at each splitting.}
\label{fig:dressing}
\end{figure}

Figure~\ref{fig:dressing}(a) shows how to read this sign for the $m=1$ pair $|1,+\rangle_b,|2,-\rangle_b$.
Back in the original frame, both branches contribute amplitudes to each laboratory $\sigma_z$ outcome.
Reversing the relative sign interchanges the two spin-resolved position densities without changing their sum.
At the marked position the partner amplitudes agree: the in-phase state gives only $\sigma_z=+1$, the out-of-phase state only $\sigma_z=-1$.

For $m>0$, tunneling makes the bare exchange vary in time.
Exchange and tunneling each flip the spin; in either order they return it to its original branch.
In the first channel, the difference between the two orderings leaves the same position-weighted reflection on both branches.
Consider then the overlap of the wavefunction with its mirror image, $\langle P_a\rangle$.
Reflection leaves the kinetic energy unchanged, while the displaced harmonic potential differs at opposite positions by an amount proportional to position.
The free rate of the mirror overlap therefore contains the same oscillator factor as the tunneling-induced rate of exchange:
\begin{equation}
 \begin{aligned}
 \dot C_{1,0}&=-i\delta(a+a^\dagger)P_a,\\
 \dot P_a&=2i\alpha(a+a^\dagger)P_a .
 \end{aligned}
 \label{eq:first_rates}
\end{equation}
Dots denote derivatives with respect to $\omega t$.
The two coefficients fix the weight of the mirror overlap in the conserved charge:
\begin{equation}
 \widetilde J_1=C_{1,0}+\frac{\delta}{2\alpha}P_a .
 \label{eq:first_charge}
\end{equation}
Since $P_a$ commutes with tunneling, no further correction is needed.
For a pair prepared with relative phase $\pi/3$, every loss of bare exchange is matched by an equal gain in weighted mirror overlap [Fig.~\ref{fig:dressing}(b)].
At its peak, this gain comes mainly from increased coherence between mirror positions near the oscillator origin (Fig.~S1~\cite{SupplementalMaterial}).

In the second channel, reflection again cancels the part of the exchange rate shared by both branches.
The remaining two-quantum term contains both position and momentum and changes sign between branches.
The free rate of reflection already supplies the position factor.
Weighting the reflection by momentum and branch sign gives $P_a(b^\dagger-b)\tau_z$, which samples the slope of the reflected wavefunction.
Its free evolution cancels the two-quantum term; a branch-dependent reflection cancels the residual displacement.
Only a transverse spin rate remains.
Precession in the bias field cancels it through a final term proportional to $\tau_x$, which commutes with tunneling.

For any integer channel $m$, the free evolution of each correction cancels the tunneling-induced change of the preceding term.
At order $\delta^k$, the oscillator polynomial in the correction $C_{m,k}$ has degree at most $m-k$.
The last term commutes with tunneling, giving the finite conserved charge $\widetilde J_m=\sum_{k=0}^{m}\delta^k C_{m,k}$~\cite{SupplementalMaterial}.
With unit coefficient for the bare exchange, this is the known minimal-degree generator~\cite{Mangazeev2021,ReyesBustos2021,XieChen2022}.

\begin{figure}[tb]
\centering
\includegraphics[width=\columnwidth]{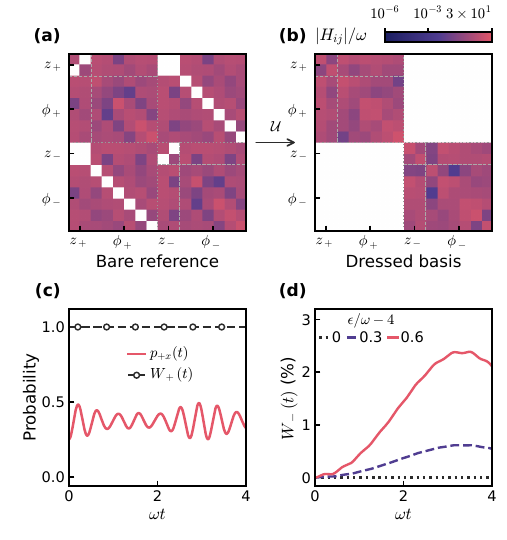}
\caption{Block diagonalization and sector dynamics at $m=4$, $g/\omega=0.7$, $\Delta/\omega=14$.
(a,b) The same Hamiltonian in the bare reference basis, $H_{\rm A}$, and in the dressed basis, $\mathcal U^\dagger H_{\rm A}\mathcal U$.
For each sign, dashes separate the two bottom ($z$) and six paired ($\phi$) states shown; dotted lines separate signs.
White denotes $|H_{ij}|/\omega<10^{-6}$.
(c) At zero detuning the dressed $z_3$ state retains $W_+=1$ while its spin occupation $p_{+x}=\langle(1+\sigma_x)/2\rangle$ oscillates.
(d) Leakage $W_-(t)$ from the same initial state into the fixed integer-bias negative sector at $d=\epsilon/\omega-4=0,0.3,0.6$.}
\label{fig:basis}
\end{figure}

At zero tunneling, the sum and difference of each resonant pair carry opposite signs [Eq.~\eqref{eq:pair_basis}].
What becomes of the $m$ unpaired states $|z_n\rangle=|n,-\rangle_b$, $0\le n<m$, whose bare charge vanishes [Fig.~\ref{fig:revival}(d)]?
For $m=1$, $C_{1,0}|z_0\rangle=0$, yet the mirror overlap of $|z_0\rangle$ is $e^{-2\alpha^2}>0$: Eq.~\eqref{eq:first_charge} gives its continuation positive charge to first order in $\delta$.
For any $m$, the first correction restricted to the bottom states is diagonal, with strictly positive magnitudes and alternating reflection phases~\cite{SupplementalMaterial}.
As $\delta\to0^+$, the highest bottom state $|z_{m-1}\rangle$ enters the positive sector and the signs alternate downward.
With these limiting assignments, the bottom states and the paired states $|\phi_{n,\pm}\rangle$ form a complete reference basis.

Could a state fall outside both sign sectors at finite tunneling?
No normalizable zero mode survives for $m=1$~\cite{KollarDutta2026}; we prove this for every $m\ge1$ at $\alpha>0$ and finite $\delta\ne0$~\cite{SupplementalMaterial}.
In the original frame the self-adjoint charge $J_m=T^\dagger\widetilde J_mT$ defines generalized parity by its sign:
\begin{equation}
 \begin{gathered}[c]
 {[H_{\rm A},J_m]}=0,\qquad P^{(m)}=\sgn J_m,\\
 \mathcal P_\pm=\frac{1\pm P^{(m)}}2 .
 \end{gathered}
 \label{eq:sectors}
\end{equation}
The absence of a zero mode gives $(P^{(m)})^2=1$; at zero bias $P^{(0)}=P$.
An arbitrary wavefunction can occupy both sectors with probabilities $W_+$ and $W_-$.

Parity fixes the subspaces; we choose a basis in each by transport from $\delta=0^+$.
At each step, convergent oscillator series give the sign projectors from the parameters without diagonalizing $H_{\rm A}$ or $J_m$~\cite{SupplementalMaterial}.
Projecting each group onto its next same-sign subspace and restoring orthonormality transports the basis unitarily~\cite{Bravyi2011}.
Undoing $T$ yields a physical basis with generally infinite Fock expansions; the parameter steps involve no time ramp.

The normalized sum and difference of the transported sign states give the dressed partners: $P^{(m)}$ interchanges them by reversing only the negative component.
For $|\phi_{1,+}\rangle$ transported to $\Delta/\omega=2.8$, the initially equal partner weights become $59.85\%$ on $|1,+\rangle_b$ and $33.70\%$ on $|2,-\rangle_b$, with $6.45\%$ on other levels [Fig.~\ref{fig:dressing}(c)].
Yet at this splitting $\mathcal P_-$ assigns zero weight to the dressed state and $8.13\%$ to the unchanged bare state [Fig.~\ref{fig:dressing}(d)].

Since $H_{\rm A}$ commutes with $P^{(m)}$, opposite signs cannot mix.
With the dressed basis vectors as columns of $\mathcal U$,
\begin{equation}
 \begin{gathered}
 \mathcal U^\dagger H_{\rm A}\mathcal U=
 \begin{pmatrix}H_+&0\\0&H_-\end{pmatrix},\\
 W_\pm(t)=\langle\Psi(t)|\mathcal P_\pm|\Psi(t)\rangle=W_\pm(0).
 \end{gathered}
 \label{eq:blocks_weights}
\end{equation}
Thus $\langle P^{(m)}\rangle=W_+-W_-$ is conserved even for a superposition of signs.
At $m=4$, $\Delta/\omega=14$, the displayed cross-block norm is $11.20\omega$ in the bare basis and below $4\times10^{-13}\omega$ in the dressed basis; couplings remain within blocks [Fig.~\ref{fig:basis}(a,b)].
For the dressed descendant of $|z_3\rangle$, spin occupation $p_{+x}(t)$ oscillates while $W_+(t)=1$ [panel (c)].
Detuning $\epsilon/\omega$ from $4$ transfers weight from the same initial state into the fixed integer-bias negative sector [panel (d)].

\begin{figure}[!tbp]
\centering
\includegraphics[width=\columnwidth]{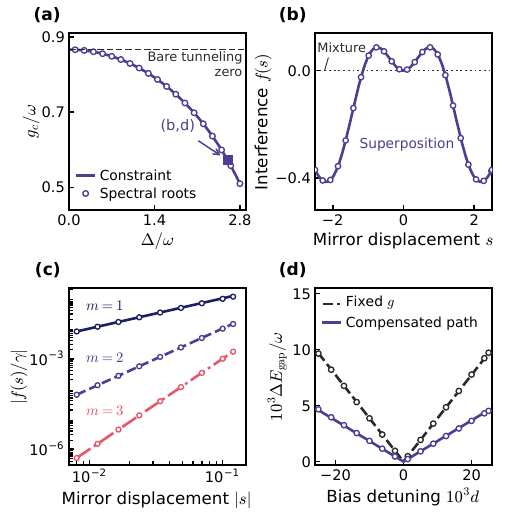}
\caption{Locating crossings and predicting their mixing from mirror interference.
(a) First-family $m=2$ crossing: constraint~\eqref{eq:first_crossing} and spectral roots (circles) versus the bare tunneling zero.
The square selects the doublet in (b,d): $\Delta/\omega=2.6$, $g_c/\omega=0.572276$.
(b) Signed signal for $(|\psi_+\rangle-i|\psi_-\rangle)/\sqrt2$ (circles: numerical; solid: analytical) and the equal incoherent mixture (dotted).
(c) First-family crossings at the same splitting: numerical $|f(s)/\gamma|$ (circles) and $|s|^m$ (lines) for $m=1,2,3$.
(d) Gaps (circles) and first-order predictions (lines) versus $d=\epsilon/\omega-2$.
At fixed $\Delta$, tuning $g(d)/\omega=g_c/\omega-0.145324d$ cancels the relative first-order shift; the gap slope measures mixing alone.}
\label{fig:response}
\end{figure}

Symmetry allows opposite-parity levels to cross; the stationary equation tells us where.
At a bare ladder energy $E/\omega=n+m/2-\alpha^2$, a finite oscillator expansion about one displaced center fixes the coefficients from the highest power down.
One bottom amplitude must vanish: this terminates the confluent Heun series~\cite{Zhong2013,Zhong2014} and selects a root of the constraint polynomial of Ku{\'s}~\cite{Kus1985} or its finite-bias extension~\cite{LiBatchelor2015,LiBatchelor2016,Kimoto2021}.
At a root with $\alpha\delta\ne0$, $J_m$ maps the resulting finite eigenstate to its degenerate opposite-center partner~\cite{SupplementalMaterial}.
The first family ($n=1$) obeys
\begin{equation}
 4\alpha_c^2+\delta^2=m+1,\qquad \delta^2<m+1 .
 \label{eq:first_crossing}
\end{equation}
The bare tunneling amplitude vanishes at $\alpha=\sqrt{m+1}/2$ regardless of splitting~\cite{LiFerriGAA2021}.
The spectral crossing instead moves to smaller coupling as tunneling increases, following Eq.~\eqref{eq:first_crossing} [Fig.~\ref{fig:response}(a)].

At an isolated finite-polynomial (Juddian) crossing, choose real orthonormal states $|\psi_\pm\rangle$ of opposite parity and energy $E$.
Their superpositions $(|\psi_+\rangle\pm i|\psi_-\rangle)/\sqrt2$ and their equal incoherent mixture all have $W_+=W_-=1/2$.
To distinguish them, we reflect the oscillator wavefunction about $X=s/\sqrt2$ using $\mathcal D(s)P_a$ for real $s$, with $X=(a+a^\dagger)/\sqrt2$~\cite{Banaszek1996,Lutterbach1997}.
The spin factor $\tau_y$ selects the phase-sensitive part of this mirror overlap:
\begin{equation}
 f(s)=\operatorname{Im}\langle\psi_+|\tau_y\mathcal D(s)P_a|\psi_-\rangle .
 \label{eq:probe}
\end{equation}
The superpositions give opposite signals $\mp f(s)$, whereas the mixture gives zero at every displacement [Fig.~\ref{fig:response}(b)].

Why does even the coherent signal vanish at the center of Fig.~\ref{fig:response}(b)?
At zero displacement the probe in the common-center frame is $\tau_y=-\dot{\tau}_x/m$ for $m\ge1$.
A time derivative has zero matrix element between states of equal energy.
Each displacement derivative inserts $a^\dagger-a$; fewer than $m$ steps cannot connect the bare partners.
Tunneling mixes in other rungs, yet the rate balance for $J_m$ reduces nonresonant terms through order $m$ to total time derivatives.
Only resonant exchange survives at equal energy:
\begin{equation}
 f(s)=\gamma s^m[1+O(s^2)],\qquad \gamma\ne0 .
 \label{eq:central}
\end{equation}
The mirror signal cannot vanish identically: only the upper spin components contribute the highest power of its polynomial factor.
For a nonzero signal the stationary equation fixes the lowest power at the origin to $m$ and the profile up to $\gamma$~\cite{Kimoto2021,SupplementalMaterial}.
Figure~\ref{fig:response}(c) resolves this power for $m=1,2,3$.

The power counts exchanged quanta; what does the amplitude $\gamma$ predict?
Detuning $d=\epsilon/\omega-m$ adds $d\tau_z/2$ to $h$.
At fixed $g$ the diagonal elements of $\tau_z$ shift the levels and its off-diagonal element mixes them; both effects can open a gap~\cite{BatchelorLiZhou2016}.
Changing $g$ at integer bias does not mix $|\psi_+\rangle$ and $|\psi_-\rangle$ to first order~\cite{SupplementalMaterial}.
If their coupling slopes differ, tuning $g$ at fixed $\Delta$ cancels the relative shift to first order in $d$.
To find this tuning at fixed small $d$, scan $g$ near the crossing for the minimum doublet gap.
Up to a common shift, the dimensionless first-order perturbation is
\begin{equation}
 \frac d2
 \begin{pmatrix}
 0&\langle\psi_+|\tau_z|\psi_-\rangle\\
 \langle\psi_-|\tau_z|\psi_+\rangle&0
 \end{pmatrix} .
 \label{eq:compensated_doublet}
\end{equation}
Its level separation, $|d|\,|\langle\psi_+|\tau_z|\psi_-\rangle|$, now measures mixing alone.

To relate this mixing to the mirror signal, consider the charge rate $id[\tau_z,J_m]/2$ induced by detuning.
Only branch-flipping terms of $J_m$ contribute in the $\tau_z$ basis.
Those terms contain the bare $m$-quantum exchange with unit coefficient.
The remaining contributions reduce to time derivatives, leaving at equal energy precisely the coherence measured by $f^{(m)}(0)=m!\gamma$.
The two states carry charges $\pm q$, with $q^2=p_m(E/\omega)>0$ from $J_m^2=p_m(h)$~\cite{ReyesBustos2021}.
The commutator weights the $\tau_z$ transition matrix element by the charge difference, of magnitude $2q$, giving~\cite{SupplementalMaterial}
\begin{equation}
 \lim_{d\to0}\frac{\Delta E_{\mathrm{gap}}^{\mathrm{comp}}}{\omega|d|}
 =|\langle\psi_+|\tau_z|\psi_-\rangle|
 =\frac{m!|\gamma|}{q}.
 \label{eq:prediction}
\end{equation}
With bare exchange normalized to unity, $q$ is the output-to-input norm ratio when $J_m$ exchanges the finite partners.
The opposite-bias constraints have a known polynomial quotient~\cite{Kimoto2021}.
We prove its conjectured relation to the charge square~\cite{ReyesBustosWakayama2022} for every $m$~\cite{SupplementalMaterial}, extending finite-order results~\cite{Steinborn2026}.
At the crossing this quotient gives $(4\alpha^2)^m q^2$.
For the marked $m=2$ doublet, $\gamma$ and $q$ predict the compensated slope $0.18407$.
At fixed $g$ the diagonal shift adds in quadrature, giving $0.38741$.
Both predictions agree with the detuned spectra without fitting [Fig.~\ref{fig:response}(d)].

In summary, at nonzero integer bias the sign of the tunneling-dressed exchange between resonant displaced-oscillator partners replaces ordinary parity.
At nonzero coupling and tunneling the charge has no normalizable zero mode: its two sign sectors exhaust the Hilbert space, including the dressed descendants of the unpaired bottom levels.
A wavefunction can occupy both sectors, and its spin and oscillator populations can evolve while the probability in each sector stays fixed.
This separation allows levels of opposite parity to cross without hybridizing.
At an isolated crossing, the small-displacement power of the spin-resolved mirror signal counts the exchanged quanta; its leading amplitude and the charge magnitude predict how strongly a small bias detuning will mix the two states.

\begin{acknowledgments}
This work was supported by the National Natural Science Foundation of China (Grant No. 12205383) and the Hunan Provincial Natural Science Foundation (Grant No. 2024JJ6483).
\end{acknowledgments}

\bibliography{references}
\end{document}

\title{Supplemental Material for Revealing the Hidden Symmetry of the Asymmetric Quantum Rabi Model}
\author{Zi-Min Li}
\author{Zi-Cheng Liu}
\author{Long-Jie Li}
\author{Xuan Zou}
\author{Jin-Xiang Fu}
\affiliation{Institute of Quantum Physics, Hunan Key Laboratory of Nanophotonics and Devices, Hunan Key Laboratory of Super-Microstructure and Ultrafast Process, School of Physics, Central South University, Changsha 410083, China}

\author{Xilin Lu}
\author{Murray T. Batchelor}
\affiliation{Mathematical Sciences Institute, The Australian National University, Canberra, ACT 2601, Australia}
\date{\today}
\maketitle

At integer bias a level crossing returns although ordinary parity is broken.
Sections~\ref{sec:start} and \ref{sec:physical} identify what is exchanged and why its dressed value is conserved.
Section~\ref{s108_complete_basis} derives the two groups of states.
Sections~\ref{sec:constraint} and \ref{sec:mirror} then determine where their energies meet and how a measurement at the crossing predicts mixing under detuning.

\tableofcontents
\clearpage

\input{01_pair}
\input{02_charge}
\input{03_basis}
\input{04_crossing}
\input{05_readout}

\medskip
\noindent\textit{AI use.}\enspace
The authors used ChatGPT and Codex to assist with derivations, numerical checks, figure preparation, and manuscript revision, and Gemini for editorial feedback.

\bibliography{references}

%% file: 01_pair.tex
\section{The revived crossing and its resonant partners}
\label{sec:start}
\subsection{What changes when the bias reaches one oscillator spacing?}

Main-text Fig.~1(a--c) compares the low-energy spectrum before ordinary parity is broken, after it is broken, and when the bias reaches one oscillator spacing.
In the spin basis $\tau_z=\sigma_x$, $\tau_x=\sigma_z$, $\tau_y=-\sigma_y$, the dimensionless Hamiltonian is
\begin{equation}
h=\frac{H}{\omega}
=N+\alpha(a+a^\dagger)\tau_z+\frac{\epsilon}{2\omega}\tau_z+\delta\tau_x,
\qquad \alpha=\frac{g}{\omega},\quad \delta=\frac{\Delta}{2\omega},
\quad N=a^\dagger a .
\end{equation}
We take $\omega>0$, $[a,a^\dagger]=1$, and real couplings.
At $\delta=0.35$, levels 2 and 3 cross at zero bias [Fig.~1(a), counting from zero].
Levels 3 and 4 have a minimum gap of $0.09002134\omega$ at $\epsilon/\omega=0.9$ [Fig.~1(b)], yet cross at $\epsilon/\omega=1$ [Fig.~1(c)], where ordinary parity is still broken.

Section~\ref{sec:spectral_windows} gives the energy windows and cutoff checks for these spectra.
The bare spin flip connects many oscillator levels at all three biases.
To see what distinguishes the integer point, switch tunneling off and ask which of those transitions costs no energy.

\subsection{The one equal-energy partner on the opposite ladder}
\label{sec:ladders}
Write $h_m=h|_{\epsilon=m\omega}$, $m=0,1,\ldots$, and $P_a=(-1)^N$ for reflection about the original oscillator origin.
The displaced branch states are $|n_\pm,\pm\rangle=\mathcal D(\mp\alpha)|n\rangle|\pm\rangle$, where $\mathcal D(s)=\exp[s(a^\dagger-a)]$ for real $s$.
This convention follows Ref.~\cite{LiBatchelor2021}.

At $\delta=0$, completing the oscillator square in each spin branch gives
\begin{equation}
h_m(0)=
\begin{pmatrix}
(a^\dagger+\alpha)(a+\alpha)-\alpha^2+m/2&0\\
0&(a^\dagger-\alpha)(a-\alpha)-\alpha^2-m/2
\end{pmatrix}.
\label{eq:sm_two_ladders}
\end{equation}
The energies $e_n^+=n-\alpha^2+m/2$ and $e_k^-=k-\alpha^2-m/2$ coincide when $k=n+m$.
Thus an $m$-quantum oscillator change can compensate the energy of a spin flip.
The lowest $m$ states of the negative branch have no resonant partner; their assignment is addressed in Sec.~\ref{s108_complete_basis}.

To compare partners using the same oscillator states, reflect only the negative spin branch:
\begin{equation}
T=\begin{pmatrix}I&0\\0&P_a\end{pmatrix},\qquad
T=T^\dagger=T^{-1},\qquad b=a+\alpha.
\end{equation}
The identity $P_a\mathcal D(\alpha)|n\rangle=(-1)^n\mathcal D(-\alpha)|n\rangle$ brings both oscillator branches to the same center.
We write $|n\rangle_b=\mathcal D(-\alpha)|n\rangle$ and $|n,s\rangle_b=|n\rangle_b|s\rangle$, with $\tau_z|s\rangle=s|s\rangle$.
Transforming the tunneling operator at the same time gives
\begin{equation}
\begin{aligned}
\widetilde h_m&=T h_mT^\dagger=h_0+\delta V,\\
h_0&=b^\dagger b-\alpha^2+\frac m2\tau_z,\\
V&=P_a\tau_x.
\end{aligned}
\label{eq:sm_common_start}
\end{equation}
Here $h_0$ has tunneling switched off while retaining the integer bias $m/2$.
The tunneling term $V$ flips the spin and reflects the oscillator.
In particular, $P_abP_a=2\alpha-b$ fixes all its oscillator matrix elements.

With $\tau_\pm=(\tau_x\pm i\tau_y)/2$, the exchange between these partners is
\begin{equation}
C_{m,0}=b^m\tau_++(b^\dagger)^m\tau_-.
\end{equation}
Its commutation with $h_0$ is the cancellation of two energy changes:
\begin{equation}
[b^\dagger b,b^m\tau_+]=-m b^m\tau_+,
\qquad [m\tau_z/2,b^m\tau_+]=m b^m\tau_+,
\end{equation}
with the adjoint cancellation for the reverse exchange.
This is why the integer bias enters before any operator ansatz is solved.
For a normalized pair superposition $|\psi\rangle=A|n,+\rangle_b+B|n+m,-\rangle_b$, the ladder matrix elements give
\begin{equation}
\langle C_{m,0}\rangle
=2\sqrt{\frac{(n+m)!}{n!}}\operatorname{Re}(A^*B),
\qquad |A|^2+|B|^2=1.
\label{eq:sm_bare_coherence}
\end{equation}
Equal populations with relative phases zero and $\pi$ have opposite exchange expectations, although their spin populations agree.
The exchange therefore measures the coherence between the two partners.
At zero tunneling the two energies agree, preserving that coherence during evolution.

Write the common-center sign states as $|\phi_{n,s}\rangle=(|n,+\rangle_b+s|n+m,-\rangle_b)/\sqrt2$.
In the original frame the exchange is $J_m(0)=T^\dagger C_{m,0}T$.
Its sign states acquire the lower-branch reflection phase:
\begin{equation}
T^\dagger|\phi_{n,s}\rangle=\frac{|n_+,+\rangle+s(-1)^{n+m}|(n+m)_-,-\rangle}{\sqrt2},\qquad
J_m(0)T^\dagger|\phi_{n,s}\rangle=s\sqrt{\frac{(n+m)!}{n!}}T^\dagger|\phi_{n,s}\rangle,
\quad s=\pm1.
\label{s108_physical_phase}
\end{equation}
Omitting $(-1)^{n+m}$ would compare different physical superpositions in the two frames.
For $m=0$, $C_{0,0}=\tau_x$ already commutes with $V$ and transforms back to ordinary parity $P=P_a\tau_x$.

\subsection{Reading the sign of the exchange in a bare pair}
In the first pair studied in main-text Fig.~2(a), $m=1$, $n=1$, and $\alpha=0.7$.
The two preparations have the same spin occupations in the branch basis.

At zero tunneling, panel (a) compares the two relative phases of the same partners:
\begin{equation*}
\begin{gathered}
\phi_{1,\pm}=\frac{|1,+\rangle_b\pm|2,-\rangle_b}{\sqrt2},\\
\frac{C_{1,0}}{\sqrt2}|1,+\rangle_b=|2,-\rangle_b,\qquad
\frac{C_{1,0}}{\sqrt2}|2,-\rangle_b=|1,+\rangle_b.
\end{gathered}
\end{equation*}
After undoing the conditional reflection, the two partners have coordinate spinors $u(X)|+\rangle_x$ and $v(X)|-\rangle_x$, where $X=(a+a^\dagger)/\sqrt2$ and each oscillator amplitude is normalized separately.
Projection onto the sum or difference of the two spin states gives the joint densities for the two original $\sigma_z$ outcomes:
\begin{equation}
p_\pm(X,+z)=\frac14|u(X)\pm v(X)|^2,\qquad
p_\pm(X,-z)=\frac14|u(X)\mp v(X)|^2.
\end{equation}
Their sum is $(|u|^2+|v|^2)/2$ for either preparation; reversing the exchange sign swaps the two outcomes.
The marked position $X=-0.342575\ldots$ solves $u(X)=v(X)$ in the central peak.
The integrated $+z$ probabilities are $0.492569$ and $0.507431$ for $\phi_{1,+}$ and $\phi_{1,-}$.
We use original-Fock cutoff 96, checked against 64, and 2001 positions in $[-8,8]$; panel (a) displays $[-4.2,4.2]$.

The phase test in Fig.~2(a) identifies the bare exchange at zero tunneling.
At finite splitting the wavefunction also visits unequal-energy partners; the next section tracks what these transitions do to the exchange.

%% file: 02_charge.tex
\section{How tunneling dresses the conserved exchange}
\label{sec:physical}
At fixed parameters the bare exchange can change only through tunneling: it already commutes with the uncoupled ladders.
We look for the observable whose oscillator or spin evolution cancels this change.
All rates below use dimensionless time $\omega t$.
We write $J_m$ for the full charge in the original frame and $\widetilde J_m=TJ_mT^\dagger$ in the common-center frame.

\subsection{First channel: why oscillator reflection is the correction}

Start with $m=1$.
Multiplying the two spin matrices exposes the disturbance caused by tunneling:
\begin{align}
[V,C_{1,0}]&=
\begin{pmatrix}P_ab^\dagger-bP_a&0\\0&P_ab-b^\dagger P_a\end{pmatrix}
=-(a+a^\dagger)P_a,\\
[h_0,P_a]&=[\alpha(a+a^\dagger),P_a]=2\alpha(a+a^\dagger)P_a,
\qquad [V,P_a]=0.
\label{eq:sm_m1_response}
\end{align}
The first line changes both spin branches in the same way and contains oscillator reflection.
The second line identifies reflection itself as an observable whose free evolution generates exactly the same oscillator factor.
It is therefore natural to try $C_{1,0}+\delta c_1P_a$; cancellation fixes $-1+2\alpha c_1=0$.
Equivalently, both rates contain the same Hermitian factor $i(a+a^\dagger)P_a$:
\begin{equation}
\begin{aligned}
\dot C_{1,0}&=-i\delta(a+a^\dagger)P_a,\\
\dot P_a&=2i\alpha(a+a^\dagger)P_a,\\
\widetilde J_1&=C_{1,0}+\frac{\delta}{2\alpha}P_a.
\end{aligned}
\label{eq:sm_first_complete}
\end{equation}
Hermiticity follows from $P_a(a+a^\dagger)=-(a+a^\dagger)P_a$.
The correction causes no further tunneling source because $[V,P_a]=0$.
The cancellation consequently holds at every finite $\delta$.

The added expectation is a mirror overlap of the wavefunction.
For common-center spin components $\psi_{sx}(X)$ along the physical $\sigma_x$ axis, with $X=(a+a^\dagger)/\sqrt2$,
\begin{equation}
\langle P_a\rangle
=\sum_{s=\pm1}\int dX\,\psi_{sx}^*(X)\psi_{sx}(-X).
\label{eq:sm_reflection_overlap}
\end{equation}
The full integral is real; its local real part can have either sign.
Its weighted change compensates the exchange change in main-text Fig.~2(b).

\subsubsection{Where the compensating mirror overlap comes from}

Main-text Fig.~2(b) evolves the initial state $(|1,+\rangle_b+e^{i\pi/3}|2,-\rangle_b)/\sqrt2$ at $\alpha=0.7$ and fixed $\delta=0.35$, using 801 times $0\le\omega t\le16$ and cutoff 96.
The original truncated Hamiltonian, transformed by $T$, generates the evolution.
We evaluate the bare exchange, reflection correction, and their sum on the same evolved state, subtracting each initial expectation.

The separate expectations each change by as much as $0.09620$, whereas the total charge varies by $5.4\times10^{-14}$ along the trajectory.
Changing the Fock cutoff from 64 to 96 alters the individual traces by less than $9.5\times10^{-15}$.

\begin{figure}[!ht]
\centering
\includegraphics[width=3.38in]{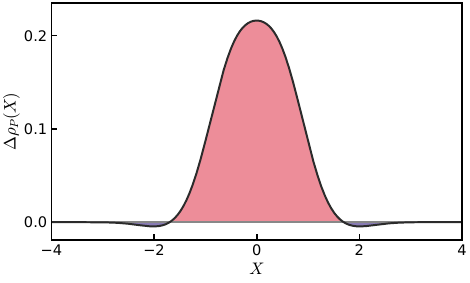}
\caption{Spatial origin of the reflection change in main-text Fig.~2(b).
At $\omega t_*=11.16$, coral and purple mark gains and losses of coherence between $X$ and $-X$ relative to the initial state.
The signed area, weighted by $\delta/(2\alpha)=0.25$, gives the reflection gain $0.09620$ and hence the magnitude of the bare exchange lost at that instant.}
\label{fig:sm_mirror_overlap}
\end{figure}

To see where the compensating reflection comes from, take the time $\omega t_*=11.16$ at which the reflection change in main-text Fig.~2(b) reaches its sampled maximum.
Undoing $T$ gives physical spin amplitudes $\psi_{sx}(X,t)=\langle X,s_x|\psi(t)\rangle$ for $s=\pm$, from which we compute
\begin{gather*}
\begin{aligned}
\rho_P(X,t)&=\operatorname{Re}\sum_{s=\pm}\psi_{sx}^*(X,t)\psi_{sx}(-X,t),\\
\Delta\rho_P(X)&=\rho_P(X,t_*)-\rho_P(X,0),
\end{aligned}\\
\frac{\delta}{2\alpha}\int\Delta\rho_P(X)\,dX
=0.25\times0.3847987018981=0.0961996754745.
\end{gather*}
The integral uses 4001 positions in $[-8,8]$; the tails outside the window in Fig.~\ref{fig:sm_mirror_overlap} contribute only $3.0\times10^{-6}$ to the weighted area.

The weighted spatial integral and the independently computed operator expectation differ by $1.7\times10^{-15}$.

\subsection{Second channel: reflected motion followed by spin precession}

For $m=2$, $C_{2,0}=b^2\tau_++(b^\dagger)^2\tau_-$ exchanges two quanta.
The same matrix multiplication gives
\begin{equation}
[V,C_{2,0}]=\bigl[\{(a^\dagger)^2-a^2\}\tau_z-2\alpha(a+a^\dagger)\bigr]P_a.
\label{eq:sm_m2_source}
\end{equation}
The displacement term is common to the branches, whereas the two-quantum term has opposite signs.
The common part is the same disturbance encountered for $m=1$, selecting $P_a$ as the first correction.
For the other part, reflection alone cannot supply the quadratic oscillator factor.
Multiplying it by momentum gives $P_a(a^\dagger-a)\tau_z$: the linear displacement in $h_0$ can now produce the required two-quantum term, with $\tau_z$ retaining the opposite branch signs.
The number-operator part also rotates momentum into displacement, leaving a lower-degree remainder.
We test this choice and then cancel that remainder using spin-dependent reflection:
\begin{align}
[h_0,P_a]&=2\alpha(a+a^\dagger)P_a,\\
[h_0,P_a(a^\dagger-a)\tau_z]&=-\bigl[2\alpha\{(a^\dagger)^2-a^2\}+a+a^\dagger\bigr]P_a\tau_z,\\
[h_0,P_a\tau_z]&=2\alpha(a+a^\dagger)P_a\tau_z.
\label{eq:sm_m2_reflected_responses}
\end{align}
Read these equations from top to bottom: the first cancels the common displacement, the second supplies the quadratic term and its linear remainder, and the third removes only that remainder.
For $C_{2,1}=c_0P_a+c_1P_a(a^\dagger-a)\tau_z+c_2P_a\tau_z$, matching the three structures requires
\begin{equation}
2\alpha c_0=2\alpha,\qquad -2\alpha c_1=-1,
\qquad -c_1+2\alpha c_2=0.
\end{equation}
The response therefore fixes the coefficients:
\begin{equation}
\begin{aligned}
C_{2,1}&=P_a+\frac{P_a(a^\dagger-a)\tau_z}{2\alpha}+\frac{P_a\tau_z}{4\alpha^2},\\
[h_0,C_{2,1}]&=-[V,C_{2,0}].
\end{aligned}
\label{eq:xterms}
\end{equation}

The momentum weighting also has a direct wavefunction meaning.
Because $a^\dagger-a=-\sqrt2\partial_X$,
\begin{equation}
\langle P_a(a^\dagger-a)\tau_z\rangle
=\sqrt2\sum_s s\int dX\,\psi_{sx}^*(X)\partial_X\psi_{sx}(-X).
\end{equation}
It reads a derivative of the reflected wavefunction with opposite branch signs.
The remaining $P_a\tau_z$ reads the difference of the two mirror overlaps.
These full products are Hermitian, even though their factors do not commute.

Tunneling acting on this correction generates the next response.
Here $[V,P_a]=[V,P_a(a^\dagger-a)\tau_z]=0$, leaving
\begin{equation}
[V,C_{2,1}]=-\frac{i\tau_y}{2\alpha^2}.
\label{eq:sm_m2_remaining_spin}
\end{equation}
This remainder is a transverse spin change with no oscillator factor.
Bias-driven precession of $\tau_x$ gives the opposite response:
\begin{equation}
\begin{aligned}
C_{2,2}&=\frac{\tau_x}{4\alpha^2},\\
[h_0,C_{2,2}]&=\frac{i\tau_y}{2\alpha^2},\\
[V,C_{2,2}]&=0.
\end{aligned}
\label{eq:sm_m2_terminal_spin}
\end{equation}
The last commutator ends the construction because this correction lies along the tunneling spin axis.
\begin{equation}
\widetilde J_2=C_{2,0}+\delta C_{2,1}+\delta^2C_{2,2},
\qquad [\widetilde h,\widetilde J_2]=0.
\end{equation}
Each component may evolve; their prescribed sum is constant at every finite $\delta$.

Returning to the original frame requires conjugating each contribution by $T$.
In particular, the common-center spin term becomes the reflected spin observable $\delta^2P_a\tau_x/(4\alpha^2)$.

\subsection{Why the corrections stop for every integer channel}
\label{sec:finite}

Both examples alternate two operations: cancel the current disturbance using free oscillator or spin motion, then calculate what tunneling does to the new term.
For general $m$, this gives the coefficient equations
\begin{align}
\widetilde J_m&=\sum_{k=0}^m\delta^kC_{m,k},\label{eq:charge_expansion}\\
[h_0,C_{m,k}]&=-[V,C_{m,k-1}],\qquad 1\le k\le m,\\
[h_0,C_{m,0}]&=[V,C_{m,m}]=0.
\label{eq:sm_balance_general}
\end{align}
The rate of each contribution makes the cancellation explicit:
\begin{equation}
\frac{d}{d(\omega t)}(\delta^kC_{m,k})
=i\delta^{k+1}[V,C_{m,k}]-i\delta^k[V,C_{m,k-1}],
\end{equation}
where the second term is absent at $k=0$ and the first vanishes at $k=m$.
The tunneling change of one contribution is canceled by the free change of the next when all rates are summed.

In both examples, tunneling converts an exchange between branches into a disturbance within each branch. The diagonal correction that cancels this disturbance can in turn generate a new exchange term. To see why this cycle ends for every $m$, we track the oscillator degree of the four spin blocks.

\subsubsection{Turning the rate balance into four block equations}

In the original branch frame, reflection changes the sign of the displacement coupling $\alpha$. Factoring $J_m=P_aM_m$ therefore turns $[h_m,J_m]=0$ into the polynomial intertwining equation $(P_ah_mP_a)M_m=M_mh_m$. The off-diagonal blocks of $M_m$ describe exchange and the diagonal blocks give its compensating response.
We take $\alpha\ne0$ and fix the upper-right block at zero tunneling to $(\alpha-a)^m$. In this convention the charge of Ref.~\cite{ReyesBustosWakayama2022} is $(2\alpha)^mJ_m$.

To keep symbol coordinates distinct from the constraint parameters $x,y$ used below, let $u,v$ be commuting Weyl-symbol variables.
The Weyl map $\W$ sends $u\mapsto-a$, $v\mapsto a^\dagger$, with completely symmetric ordering.
The number-operator commutator counts the difference between raising and lowering powers through $D=v\partial_v-u\partial_u$.
For linear and quadratic oscillator operators the symbol identities are exact:
\begin{equation}
[N,\W f]=\W(Df),\qquad [a+a^\dagger,\W f]=\W[(\partial_v+\partial_u)f],\qquad
\{a+a^\dagger,\W f\}=\W[2(v-u)f].
\end{equation}
Write the four oscillator blocks explicitly as
\begin{equation*}
M_m=\begin{pmatrix}\W d_m&\W q_m\\\W r_m&\W t_m\end{pmatrix}.
\end{equation*}
In the upper-right block, free oscillator motion and bias combine into $\mathcal E_m=D+m-\alpha(\partial_v+\partial_u)$.
The four block equations are
\begin{align}
(D-2\alpha(v-u))d_m+\delta(r_m-q_m)&=0,\\
\mathcal E_mq_m+\delta(t_m-d_m)&=0,\\
(D-m+\alpha(\partial_v+\partial_u))r_m+\delta(d_m-t_m)&=0,\\
(D+2\alpha(v-u))t_m-\delta(r_m-q_m)&=0.
\label{eq:four}
\end{align}
The diagonal blocks respond to the exchange difference $r_m-q_m$; their difference in turn feeds back into the off-diagonal exchange.
The reflection-symmetric construction uses $r_m(u,v)=q_m(v,u)$ and symmetric $d_m,t_m$, as in the first two channels.

\subsubsection{Solving the diagonal response and exchange feedback}

The exchange difference supplies the source for both diagonal blocks.
Dividing out its common antisymmetric factor defines
\begin{equation}
\mathcal A_\alpha f(u,v)=\frac{f(v,u)-f(u,v)}{2\alpha(v-u)}.
\label{eq:maps}
\end{equation}
This is a polynomial even at $u=v$, and its degree is at least one below that of $f$.
The diagonal motion requires the same division after taking the number-operator commutator:
\begin{equation}
\mathcal T_\alpha h=\frac{Dh}{2\alpha(v-u)},\qquad h(u,v)=h(v,u).
\end{equation}
For a symmetric polynomial, $Dh$ is antisymmetric and the quotient is again polynomial.
Writing such a polynomial in its sum $u+v$ and product $uv$, this operation differentiates the sum at fixed product and divides by $2\alpha$.
It therefore also lowers total degree by at least one.
Dividing the diagonal equations by $2\alpha(v-u)$ now gives
\begin{equation}
\begin{aligned}
(I-\mathcal T_\alpha)d_m&=\delta\mathcal A_\alpha q_m,\\
(I+\mathcal T_\alpha)t_m&=\delta\mathcal A_\alpha q_m.
\end{aligned}
\label{eq:sm_diagonal_response}
\end{equation}
Each repeated derivative eventually annihilates a polynomial, turning the two inverse operators into finite geometric sums.
For symmetric inputs of degree at most $m-1$, introduce the finite sums
\begin{equation}
\mathcal L_\pm=\sum_{k=0}^{m-1}(\pm\mathcal T_\alpha)^k.
\end{equation}
Thus $d_m=\delta\mathcal L_+\mathcal A_\alpha q_m$ and $t_m=\delta\mathcal L_-\mathcal A_\alpha q_m$ are the forced diagonal responses.
To return this diagonal response to the exchange block, invert its free motion on the shifted monomials:
\begin{equation}
\mathcal R_m\bigl[(v-\alpha)^i(u+\alpha)^j\bigr]
=\frac{(v-\alpha)^i(u+\alpha)^j}{m+i-j},\qquad i+j<m.
\label{eq:resolvent}
\end{equation}
Every denominator is a positive integer fixed by the mismatch of the bare oscillator transport.
Indeed, $\mathcal E_m=(v-\alpha)\partial_v-(u+\alpha)\partial_u+m$ multiplies each such monomial by $m+i-j$.
Its only kernel at degree at most $m$ is $(u+\alpha)^m$, the symbol of the resonant exchange $(\alpha-a)^m$.
Fixing its coefficient to one and inverting only the lower-degree response reduces the remaining equation to
\begin{equation}
q_m=(u+\alpha)^m+\delta^2\mathcal R_m(\mathcal L_+-\mathcal L_-)\mathcal A_\alpha q_m.
\label{eq:sm_exchange_feedback}
\end{equation}
One passage through the diagonal response and back into exchange contributes two powers of tunneling.

The finite feedback map and the four symbol blocks are
\begin{align}
\mathcal K_m&=\mathcal R_m(\mathcal L_+-\mathcal L_-)\mathcal A_\alpha,
\label{eq:feedback}\\
q_m&=\sum_{j=0}^{\lfloor m/2\rfloor}\delta^{2j}\mathcal K_m^j(u+\alpha)^m,
\nonumber\\
r_m(u,v)&=q_m(v,u),
\label{eq:finiteq}\\
d_m&=\delta\mathcal L_+\mathcal A_\alpha q_m,
\nonumber\\
t_m&=\delta\mathcal L_-\mathcal A_\alpha q_m,
\label{eq:finited}\\
J_m&=P_a\begin{pmatrix}\W d_m&\W q_m\\\W r_m&\W t_m\end{pmatrix}.
\label{eq:fullQ}
\end{align}
For $m=1$, the diagonal response is constant, making $\mathcal L_+-\mathcal L_-=0$ and leaving no exchange feedback; $m=0$ is separately $J_0=P_a\tau_x$.
Equations~\eqref{eq:maps}--\eqref{eq:fullQ} give the charge through finite coefficient operations for each input integer.
In order of use, $\mathcal A_\alpha$ forms the exchange mismatch, $\mathcal L_\pm$ solve the two diagonal responses, and $\mathcal R_m$ returns their difference to the exchange block.
The composed map $\mathcal K_m$ packages this complete correction cycle.

\subsubsection{Proof of termination and commutation}

Cancellation of the constant and all even-derivative terms in $\mathcal L_+-\mathcal L_-$ makes $\mathcal K_m$ lower degree by at least two.
Consequently $\mathcal K_m^{\lfloor m/2\rfloor+1}$ annihilates all inputs of degree at most $m$, and Eq.~\eqref{eq:finiteq} obeys exactly
\begin{equation}
(I-\delta^2\mathcal K_m)q_m=(u+\alpha)^m.
\end{equation}
The result holds for every finite $\delta$.
On the source space $(I\mp\mathcal T_\alpha)\mathcal L_\pm=I$, which proves the two diagonal equations.
The only degree-$m$ kernel of $\mathcal E_m$ is the normalized seed $(u+\alpha)^m$.
Equation~\eqref{eq:resolvent} is its explicit inverse on all lower-degree terms, and therefore
\begin{equation}
\mathcal E_mq_m=\delta^2(\mathcal L_+-\mathcal L_-)\mathcal A_\alpha q_m
=\delta(d_m-t_m).
\end{equation}
Reflection gives the other off-diagonal equation, proving all four equations in \eqref{eq:four}.
The same kernel statement gives uniqueness within the real, reflection-symmetric minimal-degree class after fixing the leading coefficient.

For the next bounds only, $q_m^{(k)}$ denotes the coefficient of $\delta^k$ in $q_m$, with the same convention for the other blocks.
The exchange blocks contain only even powers and the diagonal blocks only odd powers; their oscillator degrees obey
\begin{equation}
\deg q_m^{(2j)},\deg r_m^{(2j)}\le m-2j,\qquad
\deg d_m^{(2j+1)},\deg t_m^{(2j+1)}\le m-2j-1.
\label{eq:degree}
\end{equation}
For oscillator matrix elements the same Weyl symbol can be written in normal order:
\begin{equation}
f_{\rm normal}=e^{-\partial_u\partial_v/2}f_{\rm Weyl},\qquad
v^iu^j\mapsto(-1)^j(a^\dagger)^ia^j.
\end{equation}
For example, $\W(vu)=-N-1/2$; replacing Weyl ordering by ordinary multiplication would lose an essential constant.

For $m=1$ the diagonal blocks give the positive reflection coefficient $\delta/(2\alpha)$ of Eq.~\eqref{eq:sm_first_complete}.
For $m=2$ the only feedback in the exchange block is $\delta^2/(4\alpha^2)$, the terminal spin term in Eq.~\eqref{eq:sm_m2_terminal_spin}.
Both checks use the same normalization as the exchange derived directly from the rates~\cite{Mangazeev2021,XieChen2022}.

%% file: 03_basis.tex
\section{Why the dressed exchange divides the entire basis into two groups}
\label{s108_complete_basis}
The paired levels already have an exchange sign at zero tunneling.
The lowest $m$ oscillator states on the lower ladder have no partners: the bare exchange annihilates them.
As positive tunneling turns on, the highest of these bottom states joins the positive group and the signs alternate down the ladder [Eqs.~\eqref{s108_bottom_slope} and \eqref{s108_bottom_assignment}].
Could the dressed charge still annihilate some normalizable state at stronger tunneling?
Section~\ref{s108_no_kernel} excludes this possibility for every finite nonzero splitting.
With no states left over, the sign operator in Eq.~\eqref{s108_sign_series} determines both group weights of any bare wavefunction, and Eq.~\eqref{s108_two_blocks} gives the Hamiltonian in the resulting basis.

\subsection{From an exchange amplitude to a conserved sign}
\label{sec:parity}
At fixed parameters the expectation of the self-adjoint charge is conserved.
For a general wavefunction the corresponding two probabilities are the weights of its positive and negative charge components, not the sign of its mean charge.
We first establish why those components evolve separately.

Let $\mathcal S=\bigcap_{k\ge0}D((N+1)^k)\otimes\mathbb C^2$.
The Hamiltonian is self-adjoint on $D(N)\otimes\mathbb C^2$ and has compact resolvent because its linear oscillator term is relatively $N$-bounded with relative bound zero.
Its eigenvectors belong to $\mathcal S$: the eigenvalue equation raises oscillator regularity by one half-order at each iteration.
Polynomial oscillator operators and $P_a$ preserve $\mathcal S$.
Reflection and real coefficients make $J_m$ symmetric there, and the proven intertwining relations show that it preserves each finite-dimensional energy eigenspace.
The orthogonal direct sum of its Hermitian restrictions is a self-adjoint operator strongly commuting with $h_m$.

Finite sums of energy eigenvectors form a core for this direct sum, and the symmetry of $J_m$ identifies its action on all of $\mathcal S$ with that of the direct sum.
Thus $J_m|_{\mathcal S}$ is essentially self-adjoint, with this closure.

Generalized parity replaces every positive charge by $+1$, every negative charge by $-1$, and zero charge by zero.
The spectral theorem defines this bounded operation and its group projectors as
\begin{equation}
\begin{aligned}
P^{(m)}&=\sgn J_m,\\
\mathcal P_\pm&=\frac{I-\mathcal P_0\pm P^{(m)}}2,\\
(P^{(m)})^2&=I-\mathcal P_0.
\end{aligned}
\label{eq:sm_sign_projectors}
\end{equation}
Here $\mathcal P_0$ projects onto the kernel of the charge.
The sign removes the charge magnitude while preserving its two possible exchange orientations.
Section~\ref{s108_no_kernel} proves $\mathcal P_0=0$ for every finite $\delta\ne0$ and $\alpha>0$, reducing the physical space to exactly two sign groups.
Strictly at $\delta=0$, the $m$ unpaired low states belong to the kernel; their $0^+$ reference labels must be distinguished from this zero-tunneling charge.
For fixed parameters, strong commutation gives
\begin{equation}
\begin{gathered}
e^{ih_m\omega t}\mathcal P_\pm e^{-ih_m\omega t}=\mathcal P_\pm,\\
W_\pm(t)=\|\mathcal P_\pm\Psi(t)\|^2=W_\pm(0),
\end{gathered}
\label{eq:sm_conserved_weights}
\end{equation}
Here $t$ is physical time, as in the figures, whose horizontal axes use $\omega t$.
Each group keeps its total probability while oscillator and spin populations within it can evolve.
For a static Hamiltonian the full spectral distribution of $J_m$ is conserved.

\subsubsection{Why a possible zero charge can be tested on energy eigenstates}
\label{s108_conventions}
Here and below $m\ge1$, $\alpha>0$, and $\delta$ is finite and real.
In the common-center frame of Eq.~\eqref{eq:sm_common_start}, let $N_b=b^\dagger b$.
The complete zero-tunneling basis consists of the paired exchange states and the $m$ unpaired oscillator states:

\begingroup
\setlength{\jot}{0pt}
\setlength{\abovedisplayskip}{5pt plus 2pt minus 2pt}
\setlength{\belowdisplayskip}{5pt plus 2pt minus 2pt}
\begin{align}
&|\phi_{n,s}\rangle
=\frac{|n,+\rangle_b+s|n+m,-\rangle_b}{\sqrt2},
\qquad n\ge0,\quad s=\pm1,\nonumber\\
&C_{m,0}|\phi_{n,s}\rangle
=s\sqrt{\frac{(n+m)!}{n!}}\,|\phi_{n,s}\rangle,\nonumber\\
&|z_n\rangle=|n,-\rangle_b,\quad 0\le n<m.
\label{s108_reference_states}
\end{align}
\endgroup
The lower-spin states with $0\le n<m$ have no partner because $b^m|n\rangle_b=0$.
These pairs and bottom states jointly form a complete orthonormal basis.

Completeness of a basis is a Hilbert-space statement, while applying the charge also requires control of its large-oscillator-number matrix elements.
For a Fock expansion, normalizability requires only
\begin{equation*}
|\psi\rangle=\sum_{n,\sigma}\psi_{n,\sigma}|n,\sigma\rangle_b,
\qquad \sum_{n,\sigma}|\psi_{n,\sigma}|^2<\infty.
\end{equation*}
The bare charge multiplies paired amplitudes by a quantity growing as $n^{m/2}$, requiring the stronger weighted condition
\begin{equation*}
\sum_{\sigma=\pm}\sum_{n=0}^{\infty}(n+1)^m|\psi_{n,\sigma}|^2<\infty.
\end{equation*}
The paired eigenvalues in Eq.~\eqref{s108_reference_states} give a self-adjoint $C_{m,0}$ on
\begin{equation}
\mathcal D_m=D\bigl((N_b+\id)^{m/2}\bigr)\otimes\mathbb C^2,
\label{s108_charge_domain}
\end{equation}
with compact resolvent and an $m$-dimensional kernel.
Here compact resolvent means that the charge eigenvalues have finite multiplicity and no finite accumulation point.
The zero eigenvalue consists precisely of the bottom states.
By the degree bound~\eqref{eq:degree}, every correction in the charge expansion~\eqref{eq:charge_expansion} has oscillator degree below $m$ and hence relative bound zero with respect to $C_{m,0}$.
Explicitly, for each fixed set of parameters and every $\eta>0$, there is a finite $C_\eta$ such that
\begin{equation*}
\| (\widetilde J_m-C_{m,0})\psi\|
\le\eta\|C_{m,0}\psi\|+C_\eta\|\psi\|,
\qquad \psi\in\mathcal D_m.
\end{equation*}
The estimate follows because every smaller power of the oscillator weight is dominated by an arbitrarily small multiple of the highest power plus a constant.
The reflection $P_a$ preserves all oscillator-weighted domains because $P_abP_a=2\alpha-b$.
Consequently $\widetilde J_m$ is self-adjoint on the same domain, has compact resolvent, and is the closure identified above from the energy eigenspaces.
In particular, $J_m$ and $h_m$ strongly commute, and any nonempty kernel of $J_m$ is a finite-dimensional reducing subspace for $h_m$.
Strong commutation means that their spectral projections commute, allowing the Hamiltonian to be restricted to that kernel.
This is the reason that excluding zero charge on energy eigenvectors will suffice to exclude every physical zero-charge vector.

\subsection{How tunneling assigns the bottom states}
\label{s108_bottom_signs}

The first tunneling correction determines which of the two sectors receives each state that has no resonant partner.
We measure its slope with respect to the same tunneling parameter $\delta=\Delta/(2\omega)$ used in the Hamiltonian, keeping $\alpha$ fixed.
The projector onto the unpaired bottom states is
\begin{equation*}
\widetilde K_{\rm b}=\sum_{n=0}^{m-1}|z_n\rangle\langle z_n|.
\end{equation*}
Since the bare charge vanishes on this subspace, its first splitting is governed by the finite matrix
\begin{equation*}
\widetilde K_{\rm b}\widetilde J_m(\delta)\widetilde K_{\rm b}
=\delta\widetilde K_{\rm b}
\left.\frac{\partial\widetilde J_m}{\partial\delta}\right|_0
\widetilde K_{\rm b}+O(\delta^2).
\end{equation*}
Calculating this matrix determines how the initially zero charges split.

\subsubsection{Read the signs in the first two channels}

Before deriving the general matrix, consider the charges already constructed in Sec.~\ref{sec:physical}.
Let $c_{m,n}$ denote the diagonal slope of the bottom-state charge with respect to $\delta$.
In these two channels the restricted first-order matrix is diagonal.
For $m=1$, reflection is the only correction and there is one bottom state, $z_0$:

\begin{equation*}
c_{1,0}=\frac{e^{-2\alpha^2}}{2\alpha}>0.
\end{equation*}
The single bottom state therefore enters the positive sector for infinitesimal positive tunneling.
For $m=2$, restricting the reflection and momentum-weighted reflection terms of Sec.~\ref{sec:physical} to $z_0,z_1$ gives
\begin{equation*}
c_{2,0}=-\frac{e^{-2\alpha^2}}{4\alpha^2}<0,
\qquad
c_{2,1}=\frac{e^{-2\alpha^2}}{4\alpha^2}(1+4\alpha^2)>0.
\end{equation*}
Thus $z_0$ enters the negative sector and $z_1$ the positive sector, with generally unequal slope magnitudes.
The examples show what the general calculation must establish: which bottom states split toward positive charge and which toward negative charge.
We now compute that splitting for every $m$, including its off-diagonal matrix elements.

\subsubsection{Obtain the first correction from a finite polynomial}

The Bargmann kernel generates the Fock matrix elements of the first correction $C_{m,1}$.
For its negative-spin block we use
\begin{equation*}
\mathcal G_{m,1}(z,\xi)
=\sum_{i,j\ge0}
\frac{{}_b\langle i,-|C_{m,1}|j,-\rangle_b}{\sqrt{i!j!}}z^i\xi^j.
\end{equation*}
Thus its coefficient of $z^i\xi^j$, multiplied by $\sqrt{i!j!}$, is exactly the transition matrix element between the corresponding normalized oscillator states.
The variables are Bargmann coordinates for $b$.
In this convention the reflection kernel is $e^{-2\alpha^2+2\alpha(z+\xi)-z\xi}$: it follows by summing the overlaps of $P_a=\mathcal D(2\alpha)(-1)^{N_b}$, using $b^\dagger-b=a^\dagger-a$.
The polynomial correction therefore gives
\begin{equation}
\mathcal G_{m,1}(z,\xi)
=e^{-2\alpha^2+2\alpha(z+\xi)-z\xi}F_m(z,\xi),
\qquad \deg F_m\le m-1.
\label{s108_first_kernel}
\end{equation}
The variable $\xi$ denotes the independent antiholomorphic kernel coordinate.
The linear intertwining equation gives
\begin{equation}
\left(z\partial_z-\xi\partial_\xi+2\alpha(z-\xi)\right)F_m=z^m-\xi^m.
\label{s108_first_kernel_equation}
\end{equation}
Set $S=z+\xi$ and take $S$ derivatives at fixed product $z\xi$.
Symmetry of $F_m$ reduces this equation to
\begin{equation}
\begin{aligned}
(2\alpha+\partial_S)F_m&=\frac{z^m-\xi^m}{z-\xi},\\
F_m&=\sum_{k=0}^{m-1}\frac{(-\partial_S)^k}{(2\alpha)^{k+1}}
\frac{z^m-\xi^m}{z-\xi}.
\end{aligned}
\label{s108_first_kernel_inverse}
\end{equation}
Indeed, $z\partial_z-\xi\partial_\xi=(z-\xi)\partial_S$ at fixed $z\xi$, allowing the common factor $z-\xi$ to be removed as a polynomial identity.
The quotient is a polynomial of degree $m-1$, which its $m$th $S$ derivative annihilates.
The inverse terminates on the polynomial and imposes no smallness condition on the coupling.

\subsubsection{Extract the normalized bottom-state matrix elements}

Set $G=e^{2\alpha S}F_m$.
Equation~\eqref{s108_first_kernel_equation} becomes
\begin{equation}
(z\partial_z-\xi\partial_\xi)G
=e^{2\alpha(z+\xi)}(z^m-\xi^m).
\label{s108_diagonal_square}
\end{equation}
The right-hand side has no monomial $z^i\xi^j$ with both $i<m$ and $j<m$.
On the left-hand side, the same coefficient is $(i-j)[z^i\xi^j]G$.
Every off-diagonal coefficient of $G$ in the square $0\le i,j<m$ vanishes.
Its diagonal coefficients $d_n=[z^n\xi^n]G$ obey
\begin{align}
d_0&=\frac{(-1)^{m-1}(m-1)!}{(2\alpha)^m},
\nonumber\\
0&=\sum_{r=0}^n
\frac{d_r(-2\alpha)^{m-1+n-2r}}{(m-1-r)!(n-r)!},
\qquad 1\le n<m.
\label{s108_bottom_triangular}
\end{align}
The second relation is the vanishing $(m-1,n)$ coefficient of $F_m=e^{-2\alpha S}G$, whose total degree exceeds $m-1$.
Solving this triangular relation gives
\begin{equation}
d_n=\frac{(-1)^{m-1-n}(m-1-n)!}{n!}(2\alpha)^{2n-m}.
\label{s108_bottom_diagonal}
\end{equation}
Substitution reduces each relation to $\sum_{r=0}^n(-1)^{n-r}\binom nr=0$.
To recover a normalized Fock matrix element, multiplication by $e^{-z\xi}$ convolves these diagonal coefficients, and the Bargmann monomial normalization supplies a factor $n!$.
The resulting derivative with respect to $\delta$ is
\begin{equation*}
c_{m,n}=e^{-2\alpha^2}n!
\sum_{r=0}^n\frac{(-1)^{n-r}d_r}{(n-r)!}.
\end{equation*}
Multiplication by $e^{-z\xi}$ shifts both kernel indices equally, retaining diagonality on the bottom space.
Substituting the coefficients $d_r$ gives the normalized derivative
\begin{gather}
\widetilde K_{\rm b}\left.\frac{\partial\widetilde J_m}{\partial\delta}\right|_0\widetilde K_{\rm b}
=\sum_{n=0}^{m-1}c_{m,n}|z_n\rangle\langle z_n|,
\nonumber\\
c_{m,n}=(-1)^{m-1-n}e^{-2\alpha^2}
\sum_{r=0}^n\binom nr(m-1-r)!(2\alpha)^{2r-m}.
\label{s108_bottom_slope}
\end{gather}
Every term of the finite sum is strictly positive for $\alpha>0$, so the prefactor fixes the sign.

\subsubsection{Assemble the complete reference basis}

Positive tunneling therefore assigns
\begin{equation}
s_{m,n}=(-1)^{m-1-n},\qquad 0\le n<m.
\label{s108_bottom_assignment}
\end{equation}
The highest bottom level has positive sign, and the sign alternates on descending the lower ladder.
Enumerate the two complete reference sets as $\{|e_{j,s}\rangle\}_j$, where
\begin{equation}
\{e_{j,s}\}_j
=\{\phi_{n,s}:n\ge0\}
\cup\{z_n:0\le n<m,\ s_{m,n}=s\}.
\label{s108_complete_reference}
\end{equation}
At strictly $\delta=0$, all $z_n$ still have zero charge.
Equation~\eqref{s108_complete_reference} specifies the positive-side limiting sectors.
In particular, $(\sgn C_{m,0})^2=\id-\widetilde K_{\rm b}$ at zero tunneling.
The first-order calculation fixes the initial assignment; extending it to arbitrary finite positive tunneling requires the kernel exclusion and projector continuity proved below.

\subsection{Why no normalizable state retains zero charge}
\label{s108_no_kernel}

The bottom-state calculation assigns every reference state a sign as tunneling first turns on.
At finite tunneling the two sign groups are complete if the full charge has trivial kernel: a normalizable state satisfying $J_m\psi=0$ must have $\psi=0$.
A zero mean $\langle J_m\rangle$ can arise from cancellation between opposite charges even when $J_m\psi$ is nonzero.
Here $\alpha>0$ and $0<|\delta|<\infty$.

By Sec.~\ref{s108_conventions}, any nonempty kernel would contain an energy eigenvector. Write its energy as $e=\nu+m/2-\alpha^2$. The following tests look for a nonzero component of its image under $J_m$.
For $\nu\notin\mathbb Z_{\ge0}$, we first test the constant Taylor coefficient. Its only zeros are $\nu=-j$, $1\le j<m$, where the coefficient of order $m-j$ survives instead.
For $\nu=N\ge0$ an integer, we use the polynomial identity proved in Sec.~\ref{sec:quotient_proof}: $J_m^2$ acts on this eigenspace as $x^{-m}A_N^{(m)}(x,y)>0$, with $x=4\alpha^2$ and $y=\delta^2$. That proof starts from finite crossing states and extends the identity by polynomial interpolation.

\subsubsection{Reduce a possible zero mode to a local analytic germ}

The Bargmann representation stores the Fock amplitudes in an entire two-component function.
A single nonzero Taylor coefficient of the exchanged wavefunction establishes that the state survives the exchange.
Before conditional reflection, write the finite charge as $J_m=P_aM_m^{\rm phys}$ using Eq.~\eqref{eq:fullQ}.
In the original-oscillator Bargmann representation, set
\begin{equation}
w=z+\alpha,\qquad
\psi(z)=e^{\alpha z}\begin{pmatrix}u(w)\\v(w)\end{pmatrix},
\qquad \nu=e+\alpha^2-\frac m2,
\label{s108_local_gauge}
\end{equation}
where $e$ is the eigenvalue of $h_m$.
This invertible entire-function gauge simplifies the local differential equation while preserving whether the charge annihilates the state.
Denote the correspondingly gauged differential matrix by $M_m$.
Because both the reflection and the gauge are invertible, zero charge is equivalent to $M_m(u,v)^{\mathsf T}=0$.
The energy equation is
\begin{equation}
\begin{aligned}
wu'+(2\alpha w-\nu)u+\delta v&=0,\\
(w-2\alpha)v'-(\nu+m)v+\delta u&=0.
\end{aligned}
\label{s108_input_equations}
\end{equation}
An analytic germ here means a solution represented by a convergent Taylor series near $w=0$.
For $\nu\notin\mathbb Z_{\ge0}$, every nonzero analytic solution has $v(0)\ne0$ and can be uniquely normalized to a germ $r_\nu$ with $v(0)=1$.
Its Taylor coefficients satisfy
\begin{equation}
\begin{aligned}
u_0&=\frac\delta\nu,\quad v_0=1,\\
v_{k+1}&=\frac{\delta u_k+(k-\nu-m)v_k}{2\alpha(k+1)},\\
u_{k+1}&=\frac{2\alpha u_k+\delta v_{k+1}}{\nu-k-1}.
\end{aligned}
\label{s108_input_recurrence}
\end{equation}
None of the denominators involving $\nu$ vanishes in this case.
If $v_0$ were zero, the constant first equation would also give $u_0=0$, and the recurrence would set every subsequent coefficient to zero.
The normalization $v_0=1$ therefore loses no nonzero analytic candidate for the kernel.
Every physical Bargmann wavefunction satisfies this local analyticity condition.

The image $(U,V)^{\mathsf T}=M_mr_\nu$ obeys the energy equation with oscillator displacement reversed:
\begin{equation}
\begin{aligned}
(w-2\alpha)U'-\nu U+\delta V&=0,\\
wV'+2\alpha wV-(\nu+m)V+\delta U&=0.
\end{aligned}
\label{s108_output_equations}
\end{equation}
We call this the reflected equation below; it describes the intertwiner's output in the local gauge.
The input and output equations together will decide which coefficients of the exchanged wavefunction can vanish.

\subsubsection{Locate the obstruction in the reflected recurrence}

At $\nu=-j$, $1\le j<m$, the reflected recurrence loses one equation for a Taylor coefficient. We determine what this requires of the constant term and extract the residue needed to test the first higher coefficient.
For nonintegral $\nu$, denote the reflected analytic germ normalized to unit lower constant term by $(\widehat U,\widehat V)$.
Write $(\widehat U,\widehat V)=e^{-2\alpha w}(f,h)$, with $h_0=1$, $h_{-1}=0$.
The second reflected equation then gives
\begin{equation*}
\delta f=(\nu+m)h-wh',\qquad
f_k=\frac{\nu+m-k}{\delta}h_k.
\end{equation*}
Eliminating $f$ gives
\begin{equation}
2\alpha(k+1)(\nu+m-k-1)h_{k+1}
=[(k+x-\nu)(\nu+m-k)+y]h_k
-2\alpha(\nu+m-k+1)h_{k-1},
\label{s108_output_recurrence}
\end{equation}
where $x=4\alpha^2$ and $y=\delta^2$.
At $\nu=-j$, $1\le j<m$, set $N=\nu+m=m-j$.
The multiplier of $h_N$ vanishes, requiring the recurrence numerator at $k=N-1$ to vanish as well if $h_0$ is nonzero.
The equations for $k=0,\ldots,N-1$ form a tridiagonal system for $h_0,\ldots,h_{N-1}$.
Reversing their order gives the continuant of the fixed-$N$ recurrence
\begin{align}
P_0&=1,\qquad P_1=x+y+m-1,
\nonumber\\
P_k&=[kx+y+k(m-k)]P_{k-1}
-k(k-1)(N-k+1)xP_{k-2},
\label{s108_unmatched_constraint}
\end{align}
with $P_k=P_k^{(N,-m/2)}(x,y)$, distinct from the sector projectors.

This recurrence determinant is strictly positive.
Its symmetric continuant has diagonal entries $y+jx+j(m-j)$ and adjacent entries $\sqrt{j(j+1)(N-j)x}$, for $1\le j<N$.
Dividing row and column $j$ by $\sqrt j$ and reversing the index to $k=N-j$ gives
\begin{equation}
P_N=N!\det\left[
(a_N+\sqrt{x}\id)^\dagger(a_N+\sqrt{x}\id)
+(m-N)\id
+y\operatorname{diag}_{k=0}^{N-1}\frac1{N-k}
\right]>0.
\label{s108_positive_obstruction}
\end{equation}
Here $a_N$ is the annihilation matrix on oscillator levels $0,\ldots,N-1$.
The first term contributes a squared norm to every quadratic form, the term $(m-N)\id$ is strictly positive, and the last term is nonnegative for $y=\delta^2>0$.
No cancellation can make an eigenvalue of this auxiliary matrix zero, hence its determinant cannot vanish.
Its positive determinant establishes the recurrence obstruction.

With $h_0=1$, Eq.~\eqref{s108_output_recurrence} fixes $h_1,\ldots,h_{N-1}$ with nonzero denominators at $\nu=-j$. The numerator at step $N$ must then be nonzero: otherwise these coefficients would give a nontrivial null vector of the matrix in Eq.~\eqref{s108_positive_obstruction}. The denominator at this step is $2\alpha N(\nu+j)$, so $h_N$ has a simple pole with nonzero residue. In particular an analytic output at $\nu=-j$ must have zero lower constant term.

\subsubsection{Determine the constant conversion amplitude}

We can now test all $\nu\notin\mathbb Z_{\ge0}$ by computing the actual constant term produced by the charge.
The constant lower component of the exchanged germ is
\begin{equation}
\kappa_m(\nu):=(M_mr_\nu)_-(0)
=\frac{\delta}{(2\alpha)^m}\prod_{j=1}^{m-1}(\nu+j).
\label{s108_conversion_amplitude}
\end{equation}
The product equals one when $m=1$.

We establish Eq.~\eqref{s108_conversion_amplitude} by determining its possible poles, its degree and highest coefficient, and its roots.
The pole analysis is needed before the roots can determine the expression uniquely.

The bare lower-left entry multiplies by $w^m$ and vanishes when evaluated at $w=0$.
Its tunneling corrections start at order $\delta^2$ and have oscillator degree at most $m-2$ by Eq.~\eqref{eq:degree}.
The lower-right entry starts at order $\delta$ and has degree at most $m-1$.
Under the gauge in Eq.~\eqref{s108_local_gauge}, $a$ becomes $\partial_w+\alpha$ and $a^\dagger$ becomes $w-\alpha$; neither substitution increases derivative order.
Thus at $w=0$ these two entries require input Taylor coefficients only through orders $m-2$ and $m-1$, respectively.
For $m=1$ the lower-left contribution at this point is absent.
Equation~\eqref{s108_input_recurrence} then makes $\kappa_m$ rational in $\nu$, with possible simple poles only at $0,\ldots,m-2$.

These possible poles originate in the normalization of the input $r_\nu$.
To follow their contribution to the output, denote the second local Frobenius solution by $r_\nu^{(2)}$, normalized by its leading upper coefficient:
\begin{equation*}
(r_\nu^{(2)})_+=w^\nu\sum_{j\ge0}u_j^{(2)}w^j,\qquad
(r_\nu^{(2)})_-=w^{\nu+1}\sum_{j\ge0}v_j^{(2)}w^j.
\end{equation*}
Substituting into Eq.~\eqref{s108_input_equations} gives
\begin{equation*}
\begin{aligned}
u_0^{(2)}&=1,\qquad v_0^{(2)}=\frac{\delta}{2\alpha(\nu+1)},\\
u_j^{(2)}&=-\frac{2\alpha u_{j-1}^{(2)}+\delta v_{j-1}^{(2)}}{j},\\
v_j^{(2)}&=\frac{\delta u_j^{(2)}+(j-m)v_{j-1}^{(2)}}{2\alpha(\nu+j+1)},\qquad j\ge1.
\end{aligned}
\end{equation*}
These denominators stay nonzero near each integer $k\ge0$, making this second solution analytic in $\nu$ after removing its leading powers.
In the normalized input recurrence, the first possible pole at $\nu=k$ occurs in $u_k$; its residue has no lower coefficients and satisfies the homogeneous input equation.
Uniqueness of the recurrence with leading upper power $w^k$ therefore gives
\begin{equation*}
\operatorname*{Res}_{\nu=k}r_\nu\ \propto\ r_k^{(2)}.
\end{equation*}
The proportionality factor may be zero; the normalized input can have a pole even though the second solution is analytic in the parameter.

We next apply $M_m$ to this auxiliary solution.
Away from integer $\nu$, it has monodromy $e^{2\pi i\nu}$, the phase acquired on analytically continuing once around $w=0$.
An analytic differential intertwiner preserves this phase.
The reflected equation has exponents $0$ and $\nu+m$; the noninteger phase therefore selects the latter branch, or a zero image:
\begin{equation*}
(M_mr_\nu^{(2)})_-=O(w^{\nu+m}).
\end{equation*}
The finite differential matrix preserves the analytic parameter dependence after the leading powers are removed.
Taking $\nu\to k$ in this auxiliary image leaves no lower power than $w^{k+m}$, hence
\begin{equation*}
\operatorname*{Res}_{\nu=k}\kappa_m(\nu)
\ \propto\ (M_mr_k^{(2)})_-(0)=0.
\end{equation*}
Every possible pole of $\kappa_m$ is removable; the normalized input itself need not lose its pole.

The recurrence gives $v_k=O(\nu^k)$ and $u_k=O(\nu^{k-1})$ at fixed Taylor order.
Thus $\kappa_m$ is a polynomial of degree at most $m-1$.
The highest-degree lower diagonal block in Eq.~\eqref{eq:finited} contains the pure-annihilation term $\delta(-a)^{m-1}/(2\alpha)$.
Ordering corrections and the scalar gauge affect only lower derivative powers, fixing the highest derivative coefficient in $(M_m)_{--}$ to $(-1)^{m-1}\delta/(2\alpha)$.
The highest power in $v_k$ is $(-1)^k\nu^k/[(2\alpha)^k k!]$.
Acting with the highest derivative cancels the Taylor factorial, giving
\begin{equation*}
\frac{(-1)^{m-1}\delta}{2\alpha}(m-1)!
\frac{(-1)^{m-1}\nu^{m-1}}{(2\alpha)^{m-1}(m-1)!}
=\frac{\delta}{(2\alpha)^m}\nu^{m-1}.
\end{equation*}
All remaining derivatives contribute lower powers, fixing the highest coefficient of $\kappa_m$ to $\delta/(2\alpha)^m$.

Finally, at $\nu=-j$, $1\le j<m$, put $N=m-j$.
The output recurrence imposes the compatibility condition
\begin{equation*}
P_N^{(N,-m/2)}(x,y)\kappa_m(-j)=0.
\end{equation*}
Its strictly positive determinant~\eqref{s108_positive_obstruction} forces $\kappa_m(-j)=0$.
With the possible poles removed, these $m-1$ roots and the fixed highest coefficient prove Eq.~\eqref{s108_conversion_amplitude}.

\subsubsection{A zero constant term leaves a nonzero higher coefficient}

Let $\nu=-j$ with $1\le j<m$, and put $N=m-j$. For nearby nonintegral $\nu$, the lower output equals the normalized reflected solution times $\kappa_m(\nu)$. The input recurrence is regular at $-j$, so we may take the limit coefficient by coefficient. The coefficients below $N$ vanish because their $h_k$ stay finite while $\kappa_m(-j)=0$. At order $N$, the simple zero in Eq.~\eqref{s108_conversion_amplitude} cancels the simple pole found above, giving
\begin{equation}
[w^N](M_mr_{-j})_-=\kappa_m'(-j)\operatorname*{Res}_{\nu=-j}h_N\ne0 .
\label{s108_higher_conversion}
\end{equation}
Thus even when the constant term vanishes, the exchanged state survives. This covers the remaining negative integers at which the constant-term test fails.

\subsubsection{Close the remaining energy family and the physical kernel}

It remains to test $\nu=N=0,1,\ldots$, where the input normalization used above can fail. Here the squared norm suffices. Combining the scalar-square theorem~\cite{ReyesBustos2021} with the quotient identity~\eqref{eq:square_det} gives
\begin{equation}
J_m^2=p_m(h_m),\qquad
p_m\left(N+\frac m2-\alpha^2\right)
=x^{-m}A_N^{(m)}(x,y)>0.
\label{s108_integer_charge_square}
\end{equation}
The last inequality is the positive-quotient theorem for $x,y>0$, including $N=0$ \cite{Kimoto2021}.

For an energy eigenvector at one of these values, self-adjointness gives $\|J_m\psi\|^2=p_m(e)\|\psi\|^2>0$.
This excludes zero charge at both Juddian and non-Juddian exceptional energies in this remaining family.
Finally, a nonempty physical kernel would contain an energy eigenvector by the finite-dimensional reducing-subspace property established above.
Its Bargmann components would be entire, contradicting the exhaustive local alternatives just proved.
We conclude that
\begin{equation}
\ker J_m=\ker\widetilde J_m=\{0\},
\qquad m\ge1,\quad\alpha>0,\quad 0<|\delta|<\infty.
\label{s108_zero_kernel_theorem}
\end{equation}
The physical conclusion is that no state is left in a zero-sign subspace at finite nonzero tunneling.
The signs assigned to the bottom states can therefore enter the same complete two-group description as the resonant pairs, once continuous transport is established below.

\subsection{Projecting a bare wavefunction at any finite tunneling}
\label{s108_sign_construction}

The preceding proof establishes that the two sign groups cover the full Hilbert space.
How can we project a bare wavefunction onto these groups using the model parameters and known oscillator overlaps?
Writing $P^{(m)}=\sgn J_m$ identifies the operation; a convergent expansion makes it evaluable in the bare basis.

Exchange amplitudes grow at high oscillator excitation.
We first reduce these large amplitudes by a positive energy weighting that preserves their signs.
After rescaling the bounded result to magnitude at most one, a binomial series replaces each nonzero charge by $+1$ or $-1$.
At any finite nonzero tunneling, the resulting series for generalized parity converges on each normalizable wavefunction.

\subsubsection{Keep the inverse expansion convergent at finite tunneling}

An inverse of a positive Hamiltonian suppresses contributions from large excitation energies.
To compute this inverse without diagonalization, start from the diagonal displaced oscillator and treat each reflected spin flip as a further factor.
Their relative magnitude must remain below one even when tunneling is large.
We ensure this by shifting the energy origin by an amount chosen from the known parameters.
Start from the free Hamiltonian $h_0=N_b-\alpha^2+m\tau_z/2$ and choose
\begin{equation}
\zeta=\alpha^2+\frac m2+1+|\delta|,\qquad
\widetilde D=h_0+\zeta\id
=\operatorname{diag}(N_b+m+1+|\delta|,N_b+1+|\delta|).
\label{s108_scale}
\end{equation}
The diagonal operator $\widetilde D$ is explicitly invertible, and $\|V\|=1$ bounds the tunneling perturbation by $|\delta|$.
The shift leaves a lower bound of one even after that perturbation is included.
Changing the energy origin leaves eigenvectors and exchange signs unchanged.
Then $\widetilde h_m+\zeta\id\ge\id$, and its inverse has the explicit norm-convergent expansion
\begin{equation}
\begin{aligned}
\widetilde R&=(\widetilde h_m+\zeta\id)^{-1}
=\widetilde D^{-1}\sum_{r=0}^{\infty}
(-\delta V\widetilde D^{-1})^r,\\
\|\delta V\widetilde D^{-1}\|
&\le q_{\rm N}:=\frac{|\delta|}{1+|\delta|}<1.
\end{aligned}
\label{s108_neumann}
\end{equation}
The order of the factors follows from $\widetilde h_m+\zeta\id=(\id+\delta V\widetilde D^{-1})\widetilde D$.
Here $q_{\rm N}$ bounds the Neumann convergence ratio; it is unrelated to a charge eigenvalue.
Truncation after $r=R_0$ has norm error at most $\|\widetilde D^{-1}\|q_{\rm N}^{R_0+1}/(1-q_{\rm N})$.
All matrix elements of $V$ follow from the known displaced reflection,
\begin{equation}
{}_b\langle n|P_a|k\rangle_b
=(-1)^k\langle n|\mathcal D(2\alpha)|k\rangle,
\label{s108_reflection_elements}
\end{equation}
where $\mathcal D$ is the oscillator displacement operator.
Each term is consequently built from explicit oscillator energy denominators and reflection overlaps, with a spin flip supplied by $\tau_x$.

\subsubsection{Bound exchange amplitudes without changing their signs}

The bare exchange amplitude $\sqrt{(n+m)!/n!}$ grows as $n^{m/2}$, while the tunneling corrections have lower oscillator degree.
One factor of the positive inverse compensates one power of excitation number.
Enough inverse factors therefore bound the exchange amplitude throughout the oscillator ladder, motivating the smallest integer $k_m$ that satisfies $2k_m\ge m$.
Set $k_m=\lceil m/2\rceil$.
We use the dimensionless, bounded charge
\begin{equation}
\widetilde Y=\frac{\widetilde J_m\widetilde R^{k_m}}{L},
\label{s108_bounded_charge}
\end{equation}
where the positive scale $L$ below bounds its norm by one.
Strong commutation and positivity of $\widetilde R$ make $\widetilde Y$ self-adjoint with the same sign and kernel as $\widetilde J_m$.
On their joint spectrum the regularizing factor is $(e+\zeta)^{-k_m}>0$, which changes a charge magnitude without changing its sign or introducing a zero eigenvalue.
Expand the square polynomial about the shifted energy origin and choose the scale from its coefficients:
\begin{align*}
p_m(t-\zeta)&=\sum_{r=0}^m a_rt^r,\\
L^2&=\sum_{r=0}^m|a_r|.
\end{align*}
Since $m\le2k_m$ and $t\ge1$ on the spectrum of $\widetilde h_m+\zeta\id$,
\begin{equation}
\begin{aligned}
\widetilde Y^2&=L^{-2}p_m(\widetilde h_m)(\widetilde h_m+\zeta\id)^{-2k_m},\\
\|\widetilde Y\|&\le1.
\end{aligned}
\label{s108_parameter_bound}
\end{equation}
Indeed $t^{-2k_m}|p_m(t-\zeta)|\le\sum_r |a_r|$ for $t\ge1$.
The coefficients $a_r$ are obtained from the finite determinant in Eq.~\eqref{eq:square_det}.

\subsubsection{Recover the sign by a strongly convergent series}

Only the positive or negative sign is now needed; the bounded amplitude carries no further information for grouping states.
The scalar identity below performs precisely this normalization.
For a scalar $0<|u|\le1$, the binomial expansion gives
\begin{equation*}
u\sum_{r=0}^{\infty}4^{-r}\binom{2r}{r}(1-u^2)^r
=\frac{u}{\sqrt{1-(1-u^2)}}
=\frac{u}{|u|}.
\end{equation*}
The exclusion of $u=0$ is precisely where the physical zero-kernel theorem enters.
At a zero eigenvalue every partial sum would remain zero, leaving a state outside the two sign groups.
Applying this scalar series to the bounded charge gives the sign and projectors:
\begin{equation}
\begin{aligned}
\widetilde P^{(m)}
&=\underset{R_0\to\infty}{\operatorname{s-lim}}
\widetilde Y\sum_{r=0}^{R_0}\frac{\binom{2r}{r}}{4^r}(\id-\widetilde Y^2)^r,\\
\widetilde{\mathcal P}_s&=\frac{\id+s\widetilde P^{(m)}}{2},\qquad s=\pm1.
\end{aligned}
\label{s108_sign_series}
\end{equation}
For every nonzero scalar $u$ with $|u|\le1$, the corresponding partial sums tend to $\sgn u$ and have absolute value at most one.
This uniform bound permits dominated convergence in the spectral measure of each fixed state, even if nonzero spectral values approach zero.
The spectral theorem and the absence of a kernel therefore give strong convergence on every fixed state, together with $(\widetilde P^{(m)})^2=\id$.

The orthogonal projections $\widetilde{\mathcal P}_s$ thus sum to $\id$ even on wavefunctions with infinitely many Fock components.

For odd $m$, $k_m=(m+1)/2$ makes the regularized charge approach zero at high energy.
Resolving these small amplitudes requires increasingly many terms: the series converges on each fixed state, with truncation error depending on its oscillator tail.

The physical observables are $P^{(m)}=T^\dagger\widetilde P^{(m)}T=\sgn J_m$ and $\mathcal P_s=T^\dagger\widetilde{\mathcal P}_sT$.
In particular, $(P^{(m)})^2=\id$ at every finite nonzero tunneling covered by the theorem.
\subsubsection{Control the termwise inverse expansion}

Equations~\eqref{s108_parameter_bound} and \eqref{s108_sign_series} define the exact sign construction.
To evaluate its energy weighting term by term, we must also control the products of the inverse series with the unbounded charge.
Otherwise a small remainder of the inverse series could be amplified by the large-excitation growth of $\widetilde J_m$.
Choosing one Neumann term from each of the $k_m$ inverse factors produces the ordered term
\begin{equation*}
W_{r_1,\ldots,r_{k_m}}
=\widetilde J_m\prod_{j=1}^{k_m}
\left[\widetilde D^{-1}(-\delta V\widetilde D^{-1})^{r_j}\right],
\qquad r_j\ge0,\qquad r=\sum_{j=1}^{k_m}r_j.
\end{equation*}
Read each product from the charge on the left and cut it just after the $k_m$th occurrence of $\widetilde D^{-1}$. Each of these inverses gains one power of $N_b+\id$, while $V$ preserves the weighted oscillator domains by $P_abP_a=2\alpha-b$. Since $k_m\ge m/2$, this initial segment compensates the growth of $\widetilde J_m$. It contains at most $k_m-1$ spin flips, so the finitely many possible initial segments have a common norm bound $C$.

The remaining factors include at least $\max(r-k_m+1,0)$ complete pairs $-\delta V\widetilde D^{-1}$. Each has norm at most $q_{\rm N}<1$ by Eq.~\eqref{s108_neumann}, and every extra $\widetilde D^{-1}$ has norm at most one. Counting the ways to distribute the total order $r$ among the $k_m$ inverse factors gives
\begin{equation*}
\sum_{r_1+\cdots+r_{k_m}=r}\|W_{r_1,\ldots,r_{k_m}}\|
\le C\binom{r+k_m-1}{k_m-1}q_{\rm N}^{\max(r-k_m+1,0)}.
\end{equation*}
The binomial factor grows polynomially in $r$, so summing the geometric bound proves norm convergence for $\widetilde J_m\widetilde R^{k_m}$.

This construction supplies the two projectors at each chosen parameter value.
Following a complete basis as tunneling varies requires the separate parameter-continuity argument in the next subsection.

\subsection{Transporting the basis and reducing the Hamiltonian}
\label{s108_transport}

Transporting the entire reference basis preserves its completeness when tunneling spreads each vector over many oscillator levels.
The charge fixes the two destination subspaces, each with freedom to choose an orthonormal basis.
We choose a basis by the sequence of direct rotations defined below; changing that sequence can rotate states within a group without changing the group itself.
At each small change of tunneling, we project the old basis into the new groups and correct the change in vector lengths.
We first describe this rotation for neighboring projectors whose norm distance is less than one, and then prove that a finite sequence of such steps reaches any finite positive tunneling.
At the zero endpoint, $\widetilde{\mathcal P}_s(0^+)$ denotes the projector onto the assigned reference set in Eq.~\eqref{s108_complete_reference}; it includes the bottom states.

\subsubsection{Normalize the map between two nearby sectors}

For a prescribed finite $\delta_*>0$, choose a partition satisfying
\begin{equation}
0=\delta_0<\delta_1<\cdots<\delta_{N_{\rm step}}=\delta_*,
\quad Q_0=\widetilde{\mathcal P}_+(0^+),
\quad Q_i=\widetilde{\mathcal P}_+(\delta_i),
\quad \|Q_i-Q_{i-1}\|<1.
\label{s108_finite_partition}
\end{equation}
The zero endpoint means the positive-side extension, including the assigned bottom states.
The norm-continuity proof below guarantees that a finite number $N_{\rm step}$ of sufficiently small steps suffices on $[0,\delta_*]$.
The continuity argument establishes the required norm distance between exact projectors, while the series constructs their action on each state.
Section~\ref{s108_numerics} specifies the finite-matrix mesh and its separate cutoff checks.
On one step, first form
\begin{equation*}
A_i=Q_iQ_{i-1}+(\id-Q_i)(\id-Q_{i-1}).
\end{equation*}
Its first term maps the old positive subspace into the new one, and its second term does the same for the complementary subspace.
Its change of lengths and its subspace mapping follow directly from
\begin{equation*}
A_i^\dagger A_i=A_iA_i^\dagger=\id-(Q_i-Q_{i-1})^2,
\qquad Q_iA_i=A_iQ_{i-1}.
\end{equation*}
The norm condition makes $A_i^\dagger A_i$ strictly positive and invertible.
Removing its positive square root therefore gives the length-preserving direct rotation \cite{Bravyi2011},
\begin{equation}
W_i=A_i(A_i^\dagger A_i)^{-1/2}
=A_i\sum_{r=0}^{\infty}\frac{\binom{2r}{r}}{4^r}(Q_i-Q_{i-1})^{2r}.
\label{s108_direct_rotation}
\end{equation}
The same binomial series now converges in operator norm because $\|Q_i-Q_{i-1}\|<1$.
Thus $W_i$ is unitary and maps both $\operatorname{Ran}Q_{i-1}$ onto $\operatorname{Ran}Q_i$ and their orthogonal complements onto one another.

\subsubsection{Join local rotations into a complete basis transport}

Apply these rotations in the order of the tunneling increments:
\begin{equation}
\widetilde{\mathcal U}_m(\delta_*)
=W_{N_{\rm step}}\cdots W_1.
\label{s108_unitary_product}
\end{equation}
The product is unitary and transports both sign subspaces:
\begin{equation*}
\widetilde{\mathcal P}_s(\delta_*)\widetilde{\mathcal U}_m
=\widetilde{\mathcal U}_m\widetilde{\mathcal P}_s(0^+),\qquad s=\pm1.
\end{equation*}
Each factor intertwines the adjacent projectors, and multiplication cancels the intermediate subspace labels.
Each factor is computed from the model parameters by convergent series.
The gap that controls this transport is the charge gap at zero, so the construction also applies at energy crossings.

Apply this product to the reference vectors $|e_{j,s}\rangle$ of Eq.~\eqref{s108_complete_reference} and undo the conditional reflection:
\begin{equation}
|\chi_{j,s}(\delta_*)\rangle
=T^\dagger\widetilde{\mathcal U}_m(\delta_*)|e_{j,s}\rangle.
\label{s108_complete_dressed_basis}
\end{equation}
The physical vectors satisfy
\begin{equation}
\begin{aligned}
\langle\chi_{i,s}|\chi_{j,s'}\rangle&=\delta_{ij}\delta_{ss'},\\
\sum_{s=\pm1}\sum_j|\chi_{j,s}\rangle\langle\chi_{j,s}|&=\id,\\
P^{(m)}|\chi_{j,s}\rangle&=s|\chi_{j,s}\rangle,
\end{aligned}
\label{s108_orthogonal_complete}
\end{equation}
where completeness is a strong operator identity.
These states have a definite sign of the charge and can combine different charge magnitudes and energies within that group.
The normalized sum and difference of the two transported sign states are the images of $|n,+\rangle_b$ and $|n+m,-\rangle_b$, by linearity. Generalized parity interchanges these two images because their negative-sign component changes sign.
The plotted wavefunctions use the transport mesh specified in Sec.~\ref{s108_numerics}; their definite signs and the absence of intergroup coupling do not depend on that within-group basis choice.
In the fourth-channel example of Sec.~\ref{s108_numerics}, the charge expectation is evaluated in the transported bottom state $T^\dagger\widetilde{\mathcal U}_4|z_n\rangle$.
Its initial slope with respect to $\delta$ is Eq.~\eqref{s108_bottom_slope} at $m=4$; $n$ labels the seed and $j$ its position in the ordered basis.

\subsubsection{Why finitely many rotations reach every finite tunneling}

The rotation requires a norm bound on the distance between neighboring group projectors.
Compact resolvent and Eq.~\eqref{s108_zero_kernel_theorem} put the spectrum of the unregularized charge $\widetilde J_m$ a positive distance from zero at each fixed $\delta\ne0$.
At fixed $\alpha>0$, the charge has a common domain and polynomial dependence on $\delta$, and $\partial_\delta\widetilde J_m$ has oscillator degree at most $m-1$.
Its charge resolvent $(\widetilde J_m-it)^{-1}$, distinct from the Hamiltonian resolvent $\widetilde R$, has norm $O(|t|^{-1})$ at large $|t|$, while its norm from the Hilbert space into the charge domain stays bounded.
Equivalence of the charge graph norm with the oscillator weight in Eq.~\eqref{s108_charge_domain}, followed by interpolation to the smaller weight, gives
\begin{equation*}
\left\|(N_b+\id)^{(m-1)/2}(\widetilde J_m-it)^{-1}\right\|
=O(|t|^{-1/m}).
\end{equation*}
This weight controls $\partial_\delta\widetilde J_m$; multiplying by the other resolvent yields, locally uniformly in $\delta\ne0$,
\begin{equation}
\left\|
(\widetilde J_m-it)^{-1}
(\partial_\delta\widetilde J_m)
(\widetilde J_m-it)^{-1}
\right\|
=O(|t|^{-1-1/m})\qquad (|t|\to\infty).
\label{s108_projector_continuity}
\end{equation}
The local charge gap bounds this integrand near $t=0$.
Differentiating the resolvent representation of the sign gives
\begin{equation*}
\partial_\delta\widetilde P^{(m)}
=-\frac1\pi\int_{-\infty}^{\infty}
(\widetilde J_m-it)^{-1}(\partial_\delta\widetilde J_m)
(\widetilde J_m-it)^{-1}\,dt.
\end{equation*}
The large-$|t|$ estimate and the local gap make this derivative norm integrable.
Consequently the sector projectors are norm continuous and locally differentiable on each nonzero-tunneling interval.
This argument concerns parameter dependence of the exact sign, independently of the convergence topology of its series approximation.

At $\delta=0$, the nonzero spectrum of $C_{m,0}$ is separated from zero.
The isolated $m$-dimensional zero cluster splits according to the invertible diagonal first-order matrix in Eq.~\eqref{s108_bottom_slope}.
The projectors consequently extend in norm to $0^+$, with ranges spanned by Eq.~\eqref{s108_complete_reference}.
These are precisely the positive-side projectors used in Eq.~\eqref{s108_finite_partition}.
Their ranges include all assigned bottom states.

Norm continuity on the compact interval $[0,\delta_*]$ now supplies the finite partition required by the construction.
The zero endpoint uses the one-sided extension, and every later step stays at nonzero tunneling.

\subsubsection{Preserve the Hamiltonian domain and separate the dynamics}

Unitarity alone proves orthonormality and completeness; using the transported vectors in the Hamiltonian also requires domain preservation.
The basis vectors remain in the Hamiltonian domain.
Indeed, all Hamiltonians along this path differ by the bounded term $(\delta_i-\delta_j)V$.
Their common domain is $D(N_b)\otimes\mathbb C^2$, which contains every reference seed.
Each $Q_i$ therefore has a bounded commutator with the endpoint Hamiltonian, and $Q_0$ commutes with $h_0$.
More explicitly, strong commutation at $\delta_i$ gives
\begin{equation*}
[\widetilde h_m(\delta_*),Q_i]
=(\delta_*-\delta_i)[V,Q_i],
\end{equation*}
with the same identity at $i=0$ using $\widetilde{\mathcal P}_+(0^+)$.
The right-hand side is bounded because both $V$ and $Q_i$ are bounded.
The norm-convergent analytic normalization in Eq.~\eqref{s108_direct_rotation} preserves this bounded-commutator property and $D(\widetilde h_m)$.
Commuting the Hamiltonian through a power in that series adds at most a factor proportional to its order, which remains summable when $\|Q_i-Q_{i-1}\|<1$.
Each direct rotation and their finite product therefore preserve the Hamiltonian domain.
For the plotted fourth-channel charge means this also supplies the required quadratic-form domain, since $D(|J_4|^{1/2})=D(N_b)\otimes\mathbb C^2$ by the charge graph-norm equivalence and interpolation.
Strong commutation now gives the exact operator reduction
\begin{equation}
\widetilde{\mathcal U}_m^\dagger\widetilde h_m\widetilde{\mathcal U}_m
=\begin{pmatrix}h_+&0\\0&h_-\end{pmatrix}
\quad\hbox{on}\quad
\operatorname{Ran}\widetilde{\mathcal P}_+(0^+)\oplus\operatorname{Ran}\widetilde{\mathcal P}_-(0^+).
\label{s108_two_blocks}
\end{equation}
Both blocks are infinite dimensional and generally have nonzero off-diagonal matrix elements internally.
For a static Hamiltonian, expand any normalized state in this basis:
\begin{equation*}
|\Psi(t)\rangle=\sum_{j,s}c_{j,s}(t)|\chi_{j,s}\rangle.
\end{equation*}
The probability in each sign group is then
\begin{equation}
W_s(t)=\langle\Psi(t)|\mathcal P_s|\Psi(t)\rangle
=\sum_j|c_{j,s}(t)|^2=W_s(0).
\label{s108_group_conservation}
\end{equation}
This is the conserved weight of Eq.~\eqref{eq:sm_conserved_weights}, expressed as a sum over the complete dressed basis.
Individual coefficients, oscillator populations, and spin populations may change within each sector.
The parameter transport selects the basis used in this fixed-Hamiltonian evolution.
Under a physical tunneling ramp, the charge rate additionally contains its explicit parameter derivative.

Negative tunneling follows from the fixed spin transformation
\begin{equation}
\begin{aligned}
\widetilde h_m(-\delta)&=\tau_z\widetilde h_m(\delta)\tau_z,\\
\widetilde J_m(-\delta)&=-\tau_z\widetilde J_m(\delta)\tau_z,\\
\widetilde{\mathcal P}_s(-\delta)&=\tau_z\widetilde{\mathcal P}_{-s}(\delta)\tau_z.
\end{aligned}
\label{s108_negative_tunneling}
\end{equation}
Thus the complete construction covers every finite nonzero real $\delta$, with bottom signs reversed on the negative side in the fixed canonical convention.
For $m=0$, the binary operator is $P^{(0)}=P=\sigma_z(-1)^{a^\dagger a}$, the ordinary parity, with no unpaired bottom-state kernel.
This zero-bias case is distinct from the zero-tunneling limit at $m\ge1$, where the bottom states have zero charge.
Negative integer bias exchanges the two displaced branches.
The canonical charge normalization used here excludes $g=0$.

\subsection{Wavefunctions, matrix elements and time evolution}
\label{s108_numerics}
The two signs have now been derived for the infinite oscillator space.
The numerical tests use the Hamiltonian and charge in the original oscillator basis to check states constructed independently by transporting the bare basis.

\subsubsection{Wavefunction dressing in the first channel: Fig.~2(c,d)}

At $\alpha=0.7$, panels (c,d) start from $\phi_{1,+}$ and apply the projector and direct-rotation construction of Sec.~\ref{s108_complete_basis} as tunneling increases.
The calculation retains 48 resonant pairs and the positive bottom seed, with a 36-level resolvent buffer and 160 equal steps from $\delta=0$ to $1.4$.
Both panels sample the bare endpoint and forty positive splittings $\Delta/\omega=0.07,0.14,\ldots,2.8$ on this mesh.

Panel (c) asks how much of the transported state still occupies each bare partner.
Let $|\chi_+\rangle$ denote the state in the original frame, so that $T|\chi_+\rangle$ is its common-center representation.
The occupations of the two partners are
\[
 \big|{}_b\langle 1,+|T|\chi_+\rangle\big|^2,\qquad
 \big|{}_b\langle 2,-|T|\chi_+\rangle\big|^2 .
\]
Equivalently, in the original frame we expand the upper spin component in $\mathcal D(-\alpha)|k\rangle$ and the lower one in $\mathcal D(+\alpha)|k\rangle$.
We take the squared amplitudes at $k=1$ on the upper ladder and $k=2$ on the lower ladder, then sum the squared amplitudes on all remaining levels for the third curve.
At $\Delta/\omega=2.8$ these three probabilities are $0.598473$, $0.337007$, and $0.064520$, respectively.

Panel (d) tests hidden parity at the same splittings.
For each positive splitting we independently diagonalize the charge in the original Fock basis with 96 oscillator levels per spin.
Its negative spectral projector gives the weights of the unchanged bare seed and its transported state:
\begin{equation*}
W_-^{\rm bare}=\big\|\mathcal P_-T^\dagger|\phi_{1,+}\rangle\big\|^2,\qquad
W_-^{\rm dressed}=\big\|\mathcal P_-|\chi_+\rangle\big\|^2 .
\end{equation*}
We evaluate these squared norms from negative-eigenspace coefficients, avoiding cancellation in $(1-\langle P^{(1)}\rangle)/2$.
The bare weights are $0.0046526730$ at $\Delta/\omega=0.7$ and $0.0812559125$ at $2.8$; Table~\ref{tab:fig2_checks} gives the largest dressed weight over the scan.
At zero splitting the paired seed has a definite exchange sign; the unpaired states have zero charge.

\begin{table}[ht]
\centering
\caption{Independent tests of the transported state in main-text Fig.~2(c,d).
State changes are measured at $\Delta/\omega=0.7,2.8$; the occupation test compares all three probabilities over the full scan.}
\label{tab:fig2_checks}

\begin{tabular}{lp{0.61\textwidth}r}
\toprule
Panel & Check & Value\\
\midrule

(c,d) & State change on increasing pairs from 32 to 48 & $<2.4\times10^{-13}$\\
(c,d) & State change on refining 80 transport steps to 160 & $<7.6\times10^{-7}$\\
(c) & Largest absolute occupation change, 80 to 160 steps & $1.46\times10^{-7}$\\
(d) & Independent sign residual at $\Delta/\omega=2.8$ & $1.13\times10^{-14}$\\
(d) & Largest dressed negative-group weight & $1.8\times10^{-29}$\\
(d) & Largest sign residual over the splitting scan & $<9.6\times10^{-14}$\\
(d) & Weight change with Fock cutoff & $<1.5\times10^{-15}$\\
\bottomrule
\end{tabular}
\end{table}

\subsubsection{Construct the fourth-channel basis for Fig.~3}

The fourth-channel example uses $m=4$, $\alpha=0.7$, and $0<\delta\le7$, corresponding to $0<\Delta/\omega\le14$.
We retain $K=64$ resonant pairs, each supplying one seed per sign, and all four bottom seeds, with 36 additional oscillator levels in the resolvent calculation.
In each reference sector, bottom states in increasing oscillator order precede the paired states in increasing $n$.
Thus the first eight positive columns originate from $z_1,z_3,\phi_{0,+},\ldots,\phi_{5,+}$ and the first eight negative columns from $z_0,z_2,\phi_{0,-},\ldots,\phi_{5,-}$.

The Neumann sum in Eq.~\eqref{s108_neumann} is evaluated by repeated doubling until its geometric tail bound is below $10^{-15}$.
Let $Y_0$ be the finite matrix of $\widetilde Y=\widetilde J_m\widetilde R^{k_m}/L$ in the common-center truncated basis.
The implementation evaluates its spectral sign by the polynomial iteration
\begin{equation}
Y_{r+1}=\frac12Y_r(3\id-Y_r^2),
\qquad \|Y_r^2-\id\|_F<3\times10^{-13}.
\label{s108_numerical_sign}
\end{equation}
For each scalar $0<|u|\le1$, this iteration preserves its sign and increases its absolute value to one.
It evaluates the sign through matrix products.
Finite-matrix antisymmetry is removed by averaging the regularized charge with its transpose before the sign calculation and after each iteration.
The Hamiltonian is evaluated independently and retains all computed matrix entries.
The direct-rotation binomial sum stops when its next term has Frobenius norm below $3\times10^{-15}$.
An initial adaptive mesh accepts neighboring finite projectors with Frobenius distance below $0.6$, halving the trial step otherwise.
Two successive bisections of its intervals give the 412-step production path through $\delta=0.007,0.014,0.028,0.07,0.14,\ldots,7$.

\subsubsection{Check signs and Hamiltonian couplings independently}

After constructing the basis, we independently assemble $H$ and $J_m$ in the original Fock representation with 144 levels per spin component.
This diagonalization gives the independent parity benchmark $P_{\rm ind}^{(m)}$ and the time evolution under the original Hamiltonian.
The original-Fock cutoff is also increased to 176, and the resolvent buffer from 36 to 54.
These are finite-cutoff tests of the construction established in Sec.~\ref{s108_complete_basis}.

For the first eight physical columns in each group, test their generalized parity and their Hamiltonian coupling to every retained opposite-sign column:
\begin{align}
r_P&=\max_{s,j\le8}
\|P_{\rm ind}^{(m)}\chi_{j,s}-s\chi_{j,s}\|,
\nonumber\\
r_H&=\max_{s=\pm1}\left[
\sum_{j=1}^{8}\sum_k
\left|\left\langle\chi_{k,-s}\left|\frac H\omega\right|\chi_{j,s}\right\rangle\right|^2
\right]^{1/2}.
\label{s108_diagnostics}
\end{align}
The sum over $k$ includes every retained opposite-sector column, giving the dimensionless Frobenius coupling norm $r_H$.
At $\Delta/\omega=14$, the window in main-text Fig.~3(a,b) has a bare-reference cross-block norm $11.2005\omega$, reduced to $3.36\times10^{-13}\omega$ after dressing.
Its positive-sector off-diagonal norm remains $12.3758\omega$.
Main-text Fig.~3 displays matrix elements of magnitude below $10^{-6}\omega$ in white; the norms include all computed entries.
Across the 103 sampled tunneling values, $r_P<10^{-10}$ and the Frobenius unitarity defect of the finite transport matrix is below $4.1\times10^{-12}$.
Increasing the resolvent buffer changes the selected projected columns by less than $5.1\times10^{-16}$ in Frobenius norm at $\Delta/\omega=2.8,14$.
At the weaker check point $\delta=0.007$, the independent spectral sign diagnostic at Fock cutoff 176 gives a residual $2.85\times10^{-10}$.

Table~\ref{s108_cutoff_table} compares oscillator cutoffs on the identical transport mesh.
The change of the selected columns from $K=48$ to 64 is below $1.3\times10^{-10}$, whereas halving the transport step changes them by at most $3.4\times10^{-5}$.
The latter comparison tests refinement of the chosen basis within the fixed endpoint sectors.
\begin{table}[ht]
\centering
\caption{Independent cutoff diagnostics at $m=4$, $g/\omega=0.7$, and $\Delta/\omega=14$.
All entries are upper bounds rounded upward from the computed values.
The same transport mesh, 36-level resolvent buffer, and 144-level original-Fock benchmark are used in each row.
The tested columns are the first eight in each sign sector; $r_H$ couples them to all retained columns of the opposite sign.}
\label{s108_cutoff_table}
\begin{tabular}{rcc}
\toprule
Resonant pairs $K$ & Sign residual $r_P$ & Opposite-sector coupling $r_H$\\
\midrule
32 & $3.15\times10^{-5}$ & $3.55\times10^{-4}$\\
48 & $1.74\times10^{-10}$ & $1.53\times10^{-9}$\\
64 & $4.7\times10^{-14}$ & $5.3\times10^{-13}$\\
\bottomrule
\end{tabular}
\end{table}

\subsubsection{Bottom-state wavefunctions and their hidden-group weights}

\begin{figure}[tbp]
\centering
\includegraphics[width=3.38in]{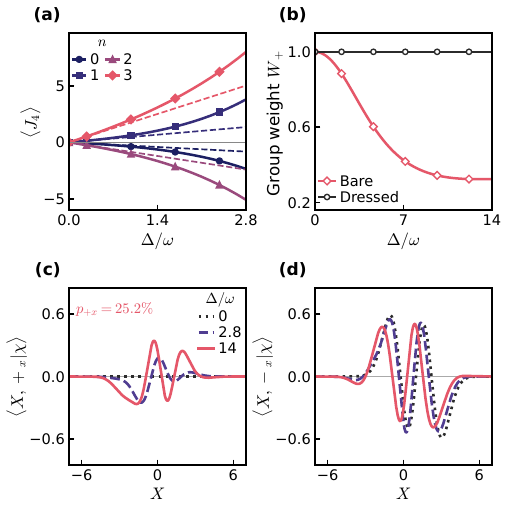}
\caption{Bottom-state assignment and wavefunction dressing at $m=4$, $\alpha=0.7$.
(a) Charge expectations for dressed bottom seeds $n=0,1,2,3$ (blue circles, indigo squares, purple triangles, coral diamonds); dashed lines are first-order limits.
(b) Positive-group weights of the dressed $z_3$ continuation (black circles) and unchanged seed (coral diamonds), with zero interpreted as the $0^+$ limit.
(c,d) Its physical positive- and negative-spin amplitudes at $\Delta/\omega=0,2.8,14$ (gray dotted, purple dashed, coral solid).
The two components share one normalization. The full state has positive hidden parity for $\Delta>0$; the bare $z_3$ has zero charge at $\Delta=0$.
The coral annotation gives the positive-spin probability at the largest splitting.}
\label{fig:sm_bottom_dressing}
\end{figure}

Physical coordinate wavefunctions are obtained after undoing $T$, including the lower-spin reflection phase in Eq.~\eqref{s108_physical_phase}.
We evaluate both spin components on 2401 equally spaced points $X=(a+a^\dagger)/\sqrt2\in[-9,9]$ with one normalization for the full two-component state.
Figure~\ref{fig:sm_bottom_dressing}(c,d) displays the transported $z_3$ at $\Delta/\omega=0,2.8,14$; its probability outside the displayed interval $|X|\le7$ is below $4.3\times10^{-12}$.
The $\Delta/\omega=14$ curves show the initial state evolved in main-text Fig.~3(c,d).
For either a bare or dressed bottom state, the assigned-sector probability is $(1+s_{m,n}\langle P^{(m)}\rangle)/2$; only the dressed state is constrained to have this probability equal to one at finite tunneling.
Figure~\ref{fig:sm_bottom_dressing}(b) shows this probability for the bare and transported $z_3$, with $s_{4,3}=+1$.

For the bottom-state plots, let $|\chi_n\rangle=T^\dagger\widetilde{\mathcal U}_4|z_n\rangle$, with $n$ labeling the bare seed.
Figure~\ref{fig:sm_bottom_dressing}(a) evaluates their charge means $q_n(\Delta)=\langle\chi_n(\Delta)|J_4(\Delta)|\chi_n(\Delta)\rangle$ at $m=4$, $\alpha=0.7$, up to $\Delta/\omega=2.8$.
The exact bare value $q_n(0)=0$ is appended to the positive-splitting samples.
The dashed first-order curves are $q_n^{(1)}=c_{4,n}\delta=c_{4,n}\Delta/(2\omega)$, using Eq.~\eqref{s108_bottom_slope} without fitting.
Their slopes with respect to $\Delta/\omega$ are $-0.293090$, $0.484575$, $-0.863716$, and $1.798317$ for $n=0,1,2,3$.
These charge expectations are evaluated in the continued basis states; the linear tangents expose departures from the weak-tunneling approximation.

The group projectors independently check definite hidden parity.
Across the 103 positive splittings used in the basis test, the 144-level original-Fock benchmark gives positive-group weights one for the continued $z_1,z_3$ and zero for the continued $z_0,z_2$, with maximum error $8.30\times10^{-13}$.
At strictly zero splitting the bare bottom states retain zero charge.

The state continued from $z_3$ remains wholly positive even when its positive-spin probability reaches $25.2\%$ at $\Delta/\omega=14$.
The original unchanged seed instead has positive-group weight $0.325$ at this parameter value.
This same dressed state supplies the dynamical test below.

\subsubsection{Spin dynamics at conserved hidden-group probability}

Main-text Fig.~3(c) evolves exactly the same dressed $z_3$ state and original Hamiltonian as Fig.~3(d), at $m=4$, $\alpha=0.7$, $\Delta/\omega=14$, and zero detuning.
It retains the same 401 time samples in $0\le\omega t\le4$.
The spin probability $p_{+x}$ is the summed probability in the original $\sigma_x=+1$ component; $W_+$ is the summed squared overlap with the constructed positive hidden basis.
Both probabilities are evaluated from the full evolved wavefunction at every time.
The spin occupation ranges from $0.251994$ to $0.492632$, whereas $W_+$ differs from one by less than $6.9\times10^{-13}$ and varies by less than $4.4\times10^{-14}$.
An independent spectral-projector evaluation agrees within $3.5\times10^{-13}$, and the negative-group probability reproduces the zero-detuning trace of Fig.~3(d).
The calculation uses 144 Fock states per spin; padding the same initial state and evolving with 176 states per spin changes the spin probability by less than $2.5\times10^{-15}$.
The basis is fixed throughout this time evolution.

For the dynamics in main-text Fig.~3(d), the initial state is the transported $z_3$ at $\Delta/\omega=14$.
We evolve it with the original Hamiltonian at fixed detunings $d=\epsilon/\omega-4=0,0.3,0.6$, retaining the same integer-bias basis for all three probability measurements.
The calculation samples 4001 times over $0\le\omega t\le40$; the displayed comparison uses $0\le\omega t\le4$.
Over the full computed interval, the maximal negative-sector probabilities are $3.1\times10^{-27}$, $0.006153$, and $0.023941$, respectively, within the finite calculation.
The sum of the two retained probabilities differs from one by less than $7.1\times10^{-13}$.
The near-zero integer-bias value is a numerical consistency check; exact conservation follows from Eq.~\eqref{s108_group_conservation}.

The vanishing intergroup matrix elements explain why levels in different groups may cross.
Their energies must still coincide; we turn next to the condition that locates the crossing.

%% file: 04_crossing.tex
\section{Where opposite signs meet and how strongly the partners exchange}
\label{sec:constraint}

The symmetry removes mixing between opposite signs; the stationary equation determines where their energies coincide.
This section derives the remaining stationary-state condition used in main-text Fig.~4(a), then relates its polynomial quotient to the exchange amplitude.
In the confluent Heun description of the unbiased and biased models, the Juddian states arise when the local series terminates~\cite{Zhong2013,Zhong2014}.
The energy alone does not imply termination: even in the unbiased model the same exceptional energy baseline can support a nonpolynomial, nondegenerate state~\cite{Maciejewski2014,Kimoto2021}.

\subsection{The last uncanceled amplitude of a finite state}

Finite wavefunctions occur at isolated parameters in Jahn--Teller systems~\cite{Judd1979} and in the unbiased Rabi model~\cite{Reik1982}.
Ku{\'s} obtained the polynomial crossing conditions for the latter~\cite{Kus1985}; Li and Batchelor treated the finite-bias constraints~\cite{LiBatchelor2015,LiBatchelor2016}.
Related derivations use polynomial truncation for several Rabi models~\cite{Moroz2018} or a mapping of the biased model to quasi-exactly solvable potentials~\cite{Guan2018}.

For fixed $n\ge1$ and $m\ge0$ put $x=4\alpha^2$ and $y=\delta^2$.
Work in the original branch frame, expanding both spin components about the positive-branch oscillator $b=a+\alpha$.
At $e_n=n+m/2-\alpha^2$, the stationary operator is
\begin{equation}
h_m-e_n=
\begin{pmatrix}
b^\dagger b-n&\delta\\
\delta&b^\dagger b-2\alpha(b+b^\dagger)+4\alpha^2-m-n
\end{pmatrix}.
\label{eq:sm_finite_stationary}
\end{equation}
Following the displaced-oscillator construction of isolated exact states~\cite{KusLewenstein1986,EmaryBishop2002}, take the finite polynomial state
\begin{equation}
L_n=\begin{pmatrix}\sum_{j=0}^nu_j(b^\dagger)^j|0_b\rangle\\
\sum_{j=0}^{n-1}v_j(b^\dagger)^j|0_b\rangle\end{pmatrix},\qquad u_n=1,
\end{equation}
The oscillator monomials here are unnormalized: $(b^\dagger)^j|0_b\rangle=\sqrt{j!}|j\rangle_b$.
This fixes the polynomial scale; the physical norm is evaluated below.
At the highest power the upper equation gives $v_n=0$, and the lower equation gives $v_{n-1}=\delta/(2\alpha)$.
The next upper equation then fixes $u_{n-1}=\delta^2/(2\alpha)$.
Each lower oscillator power supplies the next amplitude needed to cancel the stationary equation.
Encoding the downward cancellations in the normalization of Ref.~\cite{Kimoto2021} gives
\begin{align}
P_0&=1,\qquad P_1=x+y-1-m,\\
P_k&=[kx+y-k(k+m)]P_{k-1}
-k(k-1)(n-k+1)xP_{k-2}.
\label{eq:constraint_rec}
\end{align}
Here $P_k=P_k^{(n,m/2)}(x,y)$ and $F_{n,m}=P_n$; the final polynomial will supply the uncanceled bottom amplitude.
Here $P_k$ is a scalar constraint polynomial. The operator $P^{(m)}$ and its projectors $\mathcal P_\pm$ act on states.
The explicit state coefficients are
\begin{equation}
u_{n-k}=\frac{\delta^2P_{k-1}}{(2\alpha)^kk!},\qquad
v_{n-k}=\frac{\delta P_{k-1}}{(2\alpha)^k(k-1)!},\qquad 1\le k\le n.
\label{eq:statecoeff}
\end{equation}
The two spin equations at degree $j$ are
\begin{align}
(j-n)u_j+\delta v_j&=0,\\
\delta u_j+(j+4\alpha^2-m-n)v_j-2\alpha v_{j-1}-2\alpha(j+1)v_{j+1}
&=\frac{\delta P_n}{(2\alpha)^nn!}\delta_{j0}.
\end{align}
Terms outside the finite ranges vanish.
At every $j>0$, the lower equation fixes the next coefficient $v_{j-1}$, and the upper equation fixes the accompanying $u_j$.
At $j=0$ the coefficient $v_{-1}$ vanishes, so the last equation gives a condition on the model parameters.
Substitution of Eq.~\eqref{eq:statecoeff} and the recurrence proves the one-component residual used in the main text.
For example, the first family is
\begin{equation}
\begin{aligned}
&L_1=\begin{pmatrix}b^\dagger+\delta^2/(2\alpha)\\\delta/(2\alpha)\end{pmatrix}|0_b\rangle,\\
&(h_m-e_1)L_1=\frac{\delta}{2\alpha}
\bigl[4\alpha^2+\delta^2-(m+1)\bigr]|0_b,-\rangle.
\end{aligned}
\label{eq:sm_first_constraint_state}
\end{equation}
The exact crossing condition cancels this final amplitude, including the tunneling correction absent from the bare Franck--Condon zero.

\paragraph{Spectra and crossing locations in Figs.~1 and 4(a).}
\label{sec:spectral_windows}

We follow the lowest sixteen levels with 64 oscillator states per spin at 481 equally spaced couplings $0\le g/\omega\le1.6$.
At integer bias we order levels by energy within each sign of the conserved charge derived in Sec.~\ref{sec:physical}; at a degeneracy we diagonalize the charge within that energy eigenspace.
At noninteger bias levels are ordered by energy alone.
Table~\ref{tab:inset_windows} gives the enlarged windows, each sampled at 301 couplings plus the crossing or gap minimum.
At cutoffs 48 and 64 the two integer-bias gaps are below $7\times10^{-16}\omega$; the noninteger minimum is at $g/\omega=0.68910955$.
Across 41 additional couplings per bias the low energies agree to $2.0\times10^{-14}\omega$ between cutoffs; the inset energies agree to $1.2\times10^{-14}\omega$.

\begin{table}[ht]
\centering
\caption{Windows for the insets in main-text Fig.~1. Energies are indexed from zero.}
\label{tab:inset_windows}
\begin{tabular}{ccccc}
\toprule
Panel & $\epsilon/\omega$ & Energy indices & $g/\omega$ window & $E/\omega$ window\\
\midrule
(a) & $0$ & $2,3$ & $[0.35,0.59]$ & $[0.53,1.03]$\\
(b) & $0.9$ & $3,4$ & $[0.57,0.81]$ & $[0.78,1.28]$\\
(c) & $1$ & $3,4$ & $[0.57,0.81]$ & $[0.78,1.28]$\\
\bottomrule
\end{tabular}
\end{table}

Figure~4(a) uses 19 splittings $\delta\in[0.05,1.4]$ in the first family of $m=2$.
For each value, a bracket of half-width $0.035$ around the analytic coordinate is used to find a root of the independently computed signed sector energy difference.
The analytic coordinate supplies the bracket for the numerical root search.
Near degeneracy the projected charge resolves the two-state eigenspace before the energy difference is evaluated.
The same calculation is repeated at $N_{\rm cut}=48,64$.

Across these splittings the numerical roots differ from the exact coordinates by at most $4.6\times10^{-15}$ in $g/\omega$, with residual gaps below $2.2\times10^{-14}\omega$.

\subsection{Normalizing the exchanged partners and their two signs}

At a root and $\alpha\delta\ne0$, define the opposite-center partner by
\begin{equation}
L=L_n,\qquad R=P_a\tau_x L_{n+m}^{(-m/2)}.
\end{equation}
The shorter $L,R$ refer only to this fixed crossing pair; the superscript indicates that the coefficients on the right are constructed at opposite bias.
The hidden operator exchanges these monic partners \cite{ReyesBustosWakayama2022}, with the present normalization $J_mL=R$ and $J_mR=p_m(e_n)L$.
Here $J_m^2=p_m(h_m)$ is the scalar-square closure established in Ref.~\cite{ReyesBustos2021}; the next subsection determines that polynomial in the current normalization.
This normalization can also be fixed directly from the highest oscillator coefficient.
Taking norms with the self-adjoint realization described in Sec.~\ref{sec:parity} gives
\begin{equation}
\ell=\frac{L}{\|L\|},\qquad r=\frac{R}{\|R\|},\qquad
q=\frac{\|R\|}{\|L\|}=\sqrt{p_m(e_n)}>0,
\qquad J_m\ell=qr,\quad J_mr=q\ell.
\end{equation}
The real overlap $\rho=\langle\ell|r\rangle$ satisfies $|\rho|<1$ because the two finite partners are independent.
Their sign states require different normalization factors:
\begin{equation}
\begin{aligned}
&\psi_\pm=\frac{\ell\pm r}{\sqrt{2(1\pm\rho)}},\\
&\langle\psi_+|\psi_-\rangle=0,\\
&J_m\psi_\pm=\pm q\psi_\pm.
\end{aligned}
\label{eq:sm_normalized_partners}
\end{equation}
These are the real orthonormal charge states used for the mirror measurement in the main text.
Generalized parity exchanges the full partners, $P^{(m)}\ell=r$ and $P^{(m)}r=\ell$, while assigning $P^{(m)}\psi_\pm=\pm\psi_\pm$.
The partners include all tunneling corrections at the crossing.
For example, the main-text anchor has $\rho=0.9775135381$; replacing its denominators by $\sqrt2$ would be a large error.

\subsection{Identification with the polynomial quotient}

\label{sec:quotient_proof}

The divisibility relation conjectured in Ref.~\cite{Wakayama2017} and proved in Ref.~\cite{Kimoto2021} defines a polynomial $A_n^{(m)}$ before a constraint root is evaluated:
\begin{equation}
P_{n+m}^{(n+m,-m/2)}(x,y)=A_n^{(m)}(x,y)P_n^{(n,m/2)}(x,y).
\label{eq:quotient}
\end{equation}
We prove $x^mp_m(n+m/2-x/4)=A_n^{(m)}(x,y)$ using the established scalar-square theorem \cite{ReyesBustos2021} and two independent constraint identities \cite{Kimoto2021,ReyesBustosWakayama2022}.

\subsubsection{Read the exchange amplitude from the tunneling response}

At a crossing, changing the tunneling probes the spin coherence $\tau_x$ of each finite partner.
Differentiating the stationary equation expresses this response in terms of the single residual amplitude.
At a positive-$x$, nonzero-$\delta$ constraint root it gives
\begin{equation}
\langle L|\tau_x|L\rangle
=\frac{2\delta^3(\partial_yP_n)P_{n-1}}{(2\alpha)^{2n}n!(n-1)!}.
\end{equation}
To see the normalization, differentiate $(h-e_n)L=\delta P_n|0_b,-\rangle/[(2\alpha)^n n!]$ at fixed $\alpha$ and multiply by $\langle L|$ at a root.
The term containing $\partial_\delta L$ vanishes, $\partial_\delta h=\tau_x$, and $\partial_\delta P_n=2\delta\partial_yP_n$.
The remaining bottom coefficient is $v_0=\delta P_{n-1}/[(2\alpha)^n(n-1)!]$, producing the stated expression.
The fixed-family recurrence is an irreducible Jacobi problem in $y$: its real roots are simple and consecutive polynomials have no common root.
This matrix element is nonzero on the specified locus.
Differentiate the commuting relation at fixed $\alpha$, then evaluate at the original constraint root:
\begin{equation*}
[\tau_x,J_m]+[h_m,\partial_\delta J_m]=0.
\end{equation*}
Projection onto that root's energy eigenspace removes the second commutator, leaving commuting compressions of $J_m$ and $\tau_x$.
Within that doublet the two partner responses are therefore related by
\begin{equation*}
\begin{aligned}
\langle R|\tau_x|R\rangle
&=\langle L|J_m\tau_xJ_m|L\rangle\\
&=p_m(e_n)\langle L|\tau_x|L\rangle.
\end{aligned}
\end{equation*}
To compare their recurrences, write $P_k^+=P_k^{(n,m/2)}(x,y)$ and $P_k^-=P_k^{(n+m,-m/2)}(x,y)$.
The family indices $n$ and $n+m$ stay fixed while the recurrence index $k$ varies.
Their response ratio is
\begin{equation}
p_m(e_n)=\frac{\langle R|\tau_x|R\rangle}{\langle L|\tau_x|L\rangle}
=\frac{A_n^{(m)}}{x^m}
\frac{n!(n-1)!}{(n+m)!(n+m-1)!}
\frac{P_{n+m-1}^{-}}{P_{n-1}^{+}}.
\end{equation}
The established generating-coefficient identity, Theorem 7.2 of Ref.~\cite{Kimoto2021}, reads
\begin{equation}
\frac{P_k^-}{k!(k+1)!}
=\sum_{i=0}^k\binom{m}{k-i}\frac{P_i^+}{i!(i+1)!}.
\end{equation}
At $P_n^+=0$ all $P_i^+$ for $i\ge n$ vanish by the recurrence.
For $k=n+m-1$, the binomial coefficient removes all $i<n-1$, leaving just $i=n-1$.
The extra factorials cancel, proving the square--quotient identity on every such Juddian root.

\subsubsection{Extend the crossing identity using finite polynomial degree}

The physical norm ratio fixes the identity at crossings; the charge square is also needed away from them.
The finite oscillator degree makes that extension an exact polynomial interpolation problem.
Clear denominators and write
\begin{equation}
D_m(n,x,y)=x^mp_m(n+m/2-x/4)-A_n^{(m)}(x,y).
\end{equation}
We first track the coupling denominators as well as the oscillator powers in the finite construction.
For a Weyl monomial, let $s$ be its total oscillator degree, $p$ its Laurent exponent in $\alpha$, and $r$ its exponent in $\delta$.
Every term of the seed $(u+\alpha)^m$ satisfies $p+s=m$, $r=0$, and $p-s\ge-m$.
Each application of $\mathcal A_\alpha$ or $\mathcal T_\alpha$ divides by $\alpha$ and lowers the oscillator degree by one. It preserves $p-s$ and reduces $p+s$ by two. The inverse $\mathcal R_m$ divides shifted monomials by parameter-independent integers. Expanding into and out of that basis can replace $j$ oscillator factors by $\alpha^j$, preserving $p+s$ and increasing $p-s$ by $2j$.

The diagonal response $\delta\mathcal L_\pm\mathcal A_\alpha$ contains at least one degree-lowering operation and adds one power of $\delta$.
The exchange feedback $\delta^2\mathcal K_m$ contains $\mathcal A_\alpha$ and at least one $\mathcal T_\alpha$, since the constant term cancels in $\mathcal L_+-\mathcal L_-$; it adds two powers of $\delta$.
Neither operation increases $p+s+2r$, and both preserve the lower bound on $p-s$.
Induction over the finite response cycles therefore gives
\begin{equation*}
p-s\ge-m,\qquad p+s+2r\le m.
\end{equation*}
Converting to normal order only contracts pairs of oscillator factors and lowers $s$ without changing $p$ or $r$, preserving both bounds.
In particular, $s\ge0$ implies $p\ge-m$.
Thus $(2\alpha)^mJ_m$ has polynomial coefficients, and its square has weighted degree at most $4m$, with oscillator and $\alpha$ weights one and $\delta$ weight two.

Scalar-square closure expands this square in $1,h,\ldots,h^m$.
To extract the coefficient of $h^k$, compare the highest remaining normal-ordered term $(a^\dagger)^ka^k$, whose coefficient in $h^k$ is one, and then subtract that power of $h$.
Starting at $k=m$ and descending requires no division by a model parameter.
Since $h$ has weighted degree two, the coefficient of $h^k$ has weighted degree at most $4m-2k$.
Multiplying it by $h^k$ restores at most weight $4m$; subtraction therefore leaves the same bound for the next step.
Conjugations implementing $\alpha\mapsto-\alpha$ and $\delta\mapsto-\delta$ remove odd powers, making this coefficient a polynomial in $x,y$ with $x$ degree at most $2m-k$.
Substituting $e=n+m/2-x/4$ adds at most $k$ powers of $x$ and $k$ powers of $n$ to that term.
Therefore
\begin{equation*}
\deg_x\!\left[x^mp_m(n+m/2-x/4)\right]\le2m,
\qquad
\deg_n\!\left[x^mp_m(n+m/2-x/4)\right]\le m.
\end{equation*}
Independently, the established quotient determinant in Ref.~\cite{ReyesBustosWakayama2022} is
\begin{align}
A_n^{(m)}&=\det\mathcal T_m(n),\\
(\mathcal T_m)_{ii}&=y+(x-m+2i-1)(n+i),\\
(\mathcal T_m)_{i,i+1}&=n+i,\\
(\mathcal T_m)_{i+1,i}&=-i(m-i)(n+i+1).
\end{align}
Each entry is affine in $n$, giving $\deg_nD_m\le m$ and $\deg_xD_m\le2m$.

These degree bounds determine how many crossings suffice to fix the identity.
At zero tunneling, $P_n^{(n,m/2)}(x,0)=(-1)^n(n!)^2L_n^{(m)}(x)$ has $n$ distinct positive roots \cite{Kimoto2021}.
For each fixed $n>2m$ they persist for sufficiently small positive $y$.
The first part of the proof makes $D_m$ vanish at every root, more than its degree in $x$; polynomiality then extends this equality to all $x,y$ for that $n$.
Taking the $m+1$ families $n=2m+1,\ldots,3m+1$ also eliminates the polynomial dependence on $n$.
Their small-$y$ intervals may be intersected because this is a finite set.
For $m=0$, $J_0^2=A_n^{(0)}=1$ gives the result directly.

Replacing $n$ by $\nu=e+\alpha^2-m/2$ yields the finite energy formula
\begin{equation}
p_m(e)=x^{-m}\det\mathcal T_m(\nu).
\label{eq:square_det}
\end{equation}
The cleared identity extends polynomially to $x=0$; the monic charge itself remains restricted to $\alpha\ne0$.
At a constraint root, $p_m(e)$ gives the squared norm ratio of the two finite partners.

This identity plays two different roles.
Its positivity on the nonnegative integer energy family finished the zero-charge exclusion in Sec.~\ref{s108_no_kernel}.
At an actual crossing it supplies the exchange magnitude $q$; the next section uses this normalization to convert interference into a bias matrix element.

%% file: 05_readout.tex
\section{From mirror interference to the detuning gap}
\label{sec:mirror}

Having located a crossing and normalized its two sign states, we can vary their relative phase without changing their populations.
A displaced reflection reads this coherence.
We first determine its dependence on displacement, then relate its leading coefficient to the coupling that opens the crossing under bias detuning.

\subsection{What the probe measures in a crossing doublet}

In the original branch representation, take the real orthonormal states $\psi_\pm$ of Eq.~\eqref{eq:sm_normalized_partners} and the Hermitian probe $\mathcal B_s=\tau_y\mathcal D(s)P_a$.
For real $s$, its restriction to this doublet has the form
\begin{equation}
\begin{gathered}
\mathcal B_s\big|_{\rm pair}=
\begin{pmatrix}0&if(s)\\-if(s)&0\end{pmatrix},\\
\left\langle\frac{\psi_+\pm i\psi_-}{\sqrt2}\right|
\mathcal B_s
\left|\frac{\psi_+\pm i\psi_-}{\sqrt2}\right\rangle=\mp f(s).
\end{gathered}
\label{eq:sm_probe_preparation}
\end{equation}
The diagonal entries vanish because the states are real and $\tau_y$ is imaginary.
The equal incoherent mixture has the same $W_+=W_-=1/2$ as either coherent preparation, yet its probe expectation is zero.
In oscillator position, $[\mathcal D(s)P_a\psi](X)=\psi(\sqrt2s-X)$, making $s/\sqrt2$ the scanned mirror center.
The spin factor selects the relative-phase part of the two-spinor overlap.

\subsection{Equal-energy identities and the displacement equation}
\label{sec:displacement_ode}

Between equal-energy stationary states, the matrix element of any Heisenberg time derivative vanishes.
We apply this fact to the displaced reflection and its spin components, eliminating four auxiliary overlaps to obtain one equation for the measured signal.
Write the displaced reflection as $R_s=\mathcal D(s)P_a$.
For real $s$, $R_s^\dagger=R_s$, $R_s^2=I$, and
\begin{align}
R_s'&=(a^\dagger-a)R_s,\\
[N,R_s]&=s(a+a^\dagger-s)R_s,\\
[a+a^\dagger,R_s]&=2(a+a^\dagger-s)R_s,\\
\{a+a^\dagger,R_s\}&=2sR_s.
\end{align}
The measured amplitude is $f=-i\langle\psi_+|\tau_yR_s|\psi_-\rangle$ in the chosen real doublet.
Its derivatives couple to the following position-weighted overlaps:
\begin{align}
g_j&=\langle\psi_+|\tau_j(a+a^\dagger-s)R_s|\psi_-\rangle,
\qquad j=0,x,z,\quad\tau_0=I,\\
b_s&=-i\langle\psi_+|\tau_y(a+a^\dagger-s)^2R_s|\psi_-\rangle.
\end{align}
We eliminate these auxiliary overlaps using $\langle\psi_+|[h,O]|\psi_-\rangle=0$ at the common energy $e$.
The first three identities below use $O=R_s$, $\tau_zR_s$, and $\tau_xR_s$, respectively.
The fourth uses $O=-i\tau_y(a+a^\dagger-s)R_s$.
The fifth uses $\langle\psi_+|\{h,-i\tau_yR_s\}|\psi_-\rangle=2ef$ and $N=[(a+a^\dagger)^2-(a^\dagger-a)^2-2]/4$.
Writing the bias coefficient as $c=\epsilon/(2\omega)$ gives
\begin{align}
sg_0+2\alpha g_z&=0,\label{eq:mirror_g0}\\
2\alpha g_0+sg_z+2\delta f&=0,\label{eq:mirror_gz}\\
sg_x-2(\alpha s+c)f&=0,\\
f'+sb_s-2(\alpha s+c)g_x+2\delta g_z&=0,\\
4ef&=b_s-f''+(s^2-2)f-4\alpha g_x.
\end{align}
Away from $s=0,\pm2\alpha$, the first three relations fix
\begin{equation}
g_z=-\frac{2\delta s}{s^2-4\alpha^2}f,\qquad
g_0=\frac{4\alpha\delta}{s^2-4\alpha^2}f,\qquad
g_x=\frac{2(\alpha s+c)}s f.
\end{equation}
Use the fourth relation to remove $b_s$ from the fifth; this gives
\begin{equation}
f''+\frac{f'}s+
\left[4(e+\alpha^2)+2-s^2-\frac{4c^2}{s^2}
-\frac{4\delta^2}{s^2-4\alpha^2}\right]f=0.
\label{eq:mirror_ode}
\end{equation}
The temporary $g_j$ and $b_s$ are now eliminated: this single equation determines the physical overlap $f(s)$ up to its amplitude.
Eliminating $g_0$ from Eqs.~\eqref{eq:mirror_g0} and \eqref{eq:mirror_gz} gives $(s^2-4\alpha^2)g_z=-2\delta s f$. At $s=\pm2\alpha$ this forces $f=0$ for $\alpha\delta\ne0$.
At $c=m/2$, inserting a local power $f\sim s^r$ gives $r^2-m^2=0$.
Smoothness selects $r=m$ rather than the singular negative exponent, giving $f=\gamma s^m[1+O(s^2)]$; at $m=0$ the smooth branch is constant to leading order.
The central order is therefore $m$ provided $\gamma\ne0$.

\subsection{Why the leading coefficient does not vanish}

Use the finite partners $L$ and $R=P_a\tau_x L_{n+m}^{(-m/2)}$ of Sec.~\ref{sec:constraint}. On real spinors the imaginary part of the probe matrix element is antisymmetric. Changing from $(L,R)$ to $(\psi_+,\psi_-)$ thus multiplies this matrix element by a nonzero real determinant. It suffices to test the overlap between $L$ and $R$.

The reflection in the probe cancels the one in $R$. Since $\tau_y\tau_x=-i\tau_z$, the signal is proportional to the difference of the upper- and lower-component displacement overlaps between $L_n$ and $L_{n+m}^{(-m/2)}$, now at the same oscillator center.
For the unnormalized monomials of Sec.~\ref{sec:constraint}, normal ordering $\mathcal D(s)=e^{-s^2/2}e^{sb^\dagger}e^{-sb}$ gives
\begin{equation*}
e^{s^2/2}\langle0_b|b^i\mathcal D(s)(b^\dagger)^j|0_b\rangle
=(-1)^js^{i+j}+\text{lower powers of }s .
\end{equation*}
The two upper components are monic of degrees $n$ and $n+m$. Their overlap therefore supplies a nonzero term of degree $2n+m$ in $e^{s^2/2}f(s)$. The lower components reach only degrees $n-1$ and $n+m-1$, so their overlap cannot cancel this term.

The signal is consequently a Gaussian times a nonzero polynomial. If its first nonzero Taylor term at the origin is $\gamma s^k$, analyticity gives $k\ge0$ and Eq.~\eqref{eq:mirror_ode} gives $k^2=m^2$. Hence $k=m$, including $k=0$ at zero bias. Reversing the sign of either real charge state reverses $\gamma$.

\subsection{Constraint generating function and normalization}

The oscillator envelope and the central exchange order can be removed before solving for the rest of the signal.
At $e=e_n$ substitute $z=s^2$ and $f=e^{-s^2/2}s^m r(z)$.
Equation~\eqref{eq:mirror_ode} becomes
\begin{equation}
(z-x)[zr''+(m+1-z)r'+nr]-yr=0.
\end{equation}
A nonzero polynomial solution has degree $n$ and $r(x)=0$ for $y>0$.
Factoring out that side zero, write $r(z)=(z-x)u(w)$ with $w=z-x$; then
\begin{equation}
w(w+x)u''+[w(m+3-x-w)+2x]u'
+[(n-1)w+m+1-x-y]u=0.
\end{equation}
Its tridiagonal coefficient recurrence is Eq.~\eqref{eq:constraint_rec} after the normalization $k!(k+1)!$.
Explicitly, writing $u(w)=\sum_{k=0}^{n-1}a_kw^k$ gives
\begin{equation}
x(k+1)(k+2)a_{k+1}
\! =\! [(k+1)x+y-(k+1)(k+m+1)]a_k-(n-k)a_{k-1},
\end{equation}
with $a_{-1}=0$.
Thus $a_k/a_0=P_k/[x^k k!(k+1)!]$, and termination at $a_n=0$ is exactly $P_n=0$.
Equivalently, with the known generating polynomial \cite{Kimoto2021}
\begin{equation}
\Phi_{n,m}(t)=\sum_{k=0}^{n-1}\frac{P_k^{(n,m/2)}(x,y)}{k!(k+1)!}t^k,
\end{equation}
this polynomial solution and the physical overlap satisfy the same displacement equation and the same unique smooth branch at the origin.
Matching the leading coefficient $\gamma$ therefore fixes the physical signal:
\begin{equation}
f(s)=\frac{\gamma}{-x\Phi_{n,m}(-1)}
e^{-s^2/2}s^m(s^2-x)\Phi_{n,m}\!\left(\frac{s^2-x}{x}\right).
\label{eq:fullmirror}
\end{equation}
This is the constraint generating-function equation of Ref.~\cite{Kimoto2021}.
The physical identification supplies its normalized interference amplitude.
The nonzero leading coefficient ensures $\Phi_{n,m}(-1)\ne0$ on the genuine doublet.
For the first family, $n=1$ and $\Phi_{1,m}=1$.
The signed curve tested in main-text Fig.~4(b) is therefore $f(s)=\gamma e^{-s^2/2}s^m(1-s^2/x)$.

\subsection{Equal-energy selection and bias response}

The displacement equation in Sec.~\ref{sec:displacement_ode} fixes the shape of the mirror signal up to its amplitude. Which part of the exchange fixes $\gamma=f^{(m)}(0)/m!$? Expanding the probe in displacement, we show that all currents of order below $m$ are total time derivatives. The surviving $m$-quantum current also gives the off-diagonal bias matrix element.

\subsubsection{Identify the common exchange current}

For $m\ge1$, work in the common-center frame, retaining $a=b-\alpha$ because the physical displacement probe is defined in $a$, and write such a current as
\begin{equation}
\mathcal I[O]=i\begin{pmatrix}0&O\\-O^\dagger&0\end{pmatrix},\qquad
O=\sum_{i+j\le m}o_{ij}(a^\dagger)^i a^j,\qquad o_{ij}\in\mathbb R.
\end{equation}
Here $O$ is an oscillator polynomial and $\mathcal I[O]$ is its phase current.
For example, $\mathcal I[1]=-\tau_y=i[\widetilde h,\tau_x/m]$, since only the bias term fails to commute with $\tau_x$. Between two states of the same energy the commutator has zero matrix element, so the constant term of the probe cannot contribute.
The probe obeys $T\mathcal B_sT^\dagger=-\mathcal I[\mathcal D(s)]$, making its $k$th derivative at zero displacement the current $-\mathcal I[(a^\dagger-a)^k]$.
For $k<m$ this polynomial has no $a^m$ term; at $k=m$ its coefficient is $(-1)^m$, unchanged by normal ordering.
Between equal-energy states, every term other than this resonant contribution can be removed as a total Heisenberg rate.

Writing $\widetilde P_e$ for the full energy projection in this common-center frame, the selection rule is
\begin{equation}
\widetilde P_e\mathcal I[O]\widetilde P_e=o_{0m}\widetilde P_e\mathcal I[a^m]\widetilde P_e,
\label{eq:phase_select}
\end{equation}

The reason is the same rate balance used to construct the charge.
For the nonresonant part of $O$, we construct an operator $B_O$ such that
\begin{equation}
\mathcal I[O]-o_{0m}\mathcal I[a^m]=i[\widetilde h,B_O].
\end{equation}
The right-hand side has zero matrix element at equal energy.
The surviving $a^m$ term is therefore shared by the probe derivative and the exchange part of the charge.
We use this fact first, then give the finite construction of $B_O$.

\subsubsection{Read the bias matrix element from the charge itself}

The upper-right block of the common-center charge is
\begin{equation}
O_m=P_a\W q_mP_a=(a+\alpha)^m+\text{terms of degree at most }m-2.
\label{eq:sm_exchange_block}
\end{equation}
For $m=1$ the remainder vanishes.
Equation~\eqref{eq:degree} makes its leading $a^m$ coefficient exactly one, independently of tunneling.
The diagonal charge blocks commute with $\tau_z$, whereas the exchange blocks give $[\widetilde J_m,\tau_z]=2i\mathcal I[O_m]$.
Applying Eq.~\eqref{eq:phase_select} therefore yields the projected identity
\begin{equation}
\widetilde P_e[\widetilde J_m,\tau_z]\widetilde P_e
=2i\widetilde P_e\mathcal I[a^m]\widetilde P_e.
\label{eq:source}
\end{equation}
In the states $T\psi_\pm$, the left-hand side has matrix element $2q\langle\psi_+|\tau_z|\psi_-\rangle$ because $T$ leaves $\tau_z$ unchanged.
The resonant phase current fixes bias mixing from the integer-point charge.

Equation~\eqref{eq:source} gives $\langle T\psi_+|\mathcal I[a^m]|T\psi_-\rangle=-iq\langle\psi_+|\tau_z|\psi_-\rangle$.
Combining this with the probe derivatives identified above yields, in the monic real charge convention,
\begin{equation}
f^{(k)}(0)=0\quad(k<m),\qquad
f^{(m)}(0)=(-1)^m q\langle\psi_+|\tau_z|\psi_-\rangle.
\label{eq:gamma}
\end{equation}
This proves the amplitude relation; the preceding nonvanishing argument supplies its strict leading order.
For the normalization $J_m^{\rm lit}=(2\alpha)^mJ_m$, use $q=q_{\rm lit}/|2\alpha|^m$ in this relation.

\subsubsection{Construct the canceled nonresonant rates}

The amplitude relation above uses only the resonant coefficient of $O$.
We now verify that all remaining terms really are total time derivatives, using the same finite response operations that built $J_m$.
To use the block equations of Sec.~\ref{sec:finite}, first express $P_aOP_a$ as a Weyl symbol:
\begin{equation}
g_O(u,v)=e^{\frac12\partial_u\partial_v}
\sum_{i+j\le m}(-1)^i o_{ij}v^i u^j,
\qquad h_O=g_O-o_{0m}u^m.
\end{equation}
The exponential is a finite sum of contractions converting normal ordering to symmetric ordering.
It lowers degree and therefore leaves the $u^m$ coefficient unchanged.
The ordering and reflection conventions give $\W g_O(u,v)=P_aOP_a$ and $\W g_O(v,u)=O^\dagger$ for these real coefficients.
Shifting the monomials to powers of $v-\alpha$ and $u+\alpha$ changes only lower-degree terms; subtracting the $u^m$ term therefore also removes the resonant $(u+\alpha)^m$ coefficient from $h_O$.
Extend the bare transport inverse of Eq.~\eqref{eq:resolvent} to
\begin{equation}
\mathcal R_m^\circ\bigl[(v-\alpha)^i(u+\alpha)^j\bigr]
=\frac{(v-\alpha)^i(u+\alpha)^j}{m+i-j},
\qquad i+j\le m,\quad(i,j)\ne(0,m).
\end{equation}
Every denominator is now a positive integer; the sole resonant direction has been removed.
The source $h_O$ enters the upper-right block; the lower-left block receives its negative reflection, while the two diagonal equations retain zero source.
The diagonal response is therefore still $d_O=\delta\mathcal L_+\mathcal A_\alpha q_O$ and $t_O=\delta\mathcal L_-\mathcal A_\alpha q_O$.
Substituting it into the upper-right equation gives
\begin{align*}
\mathcal E_m q_O&=h_O+\delta(d_O-t_O),\\
q_O&=\mathcal R_m^\circ h_O+\delta^2\mathcal K_m q_O.
\end{align*}
Thus the inverse bare motion acts on the prescribed source instead of a resonant seed; each subsequent correction cycle is unchanged.
Since that cycle lowers oscillator degree, the feedback sum terminates and gives
\begin{align}
q_O&=\sum_{l=0}^{\lfloor m/2\rfloor}
\delta^{2l}\mathcal K_m^l\mathcal R_m^\circ h_O,
\nonumber\\
r_O(u,v)&=q_O(v,u),\\
d_O&=\delta\mathcal L_+\mathcal A_\alpha q_O,
\nonumber\\
t_O&=\delta\mathcal L_-\mathcal A_\alpha q_O.
\end{align}
Here $\mathcal K_m$ is one diagonal-response and exchange-feedback cycle, and $\mathcal L_\pm\mathcal A_\alpha$ recover its forced diagonal blocks.
Each feedback step lowers degree by at least two and therefore cannot return to the removed resonant direction.
Substitution into the four block equations gives residuals $(0,h_O(u,v),-h_O(v,u),0)$.
Multiplication by $iP_a$ and conjugation by $T$ convert this original-frame residual into $\mathcal I[O]-o_{0m}\mathcal I[a^m]$ by the ordering identities above.
Explicitly, $[h_m,P_aM]=P_a[(P_ah_mP_a)M-Mh_m]$ for any block matrix $M$.
Applying this identity to the four blocks just constructed and using $T=\operatorname{diag}(I,P_a)$ gives
\begin{align*}
i[\widetilde h,B_O]
&=iT P_a\begin{pmatrix}0&\W h_O(u,v)\\-\W h_O(v,u)&0\end{pmatrix}T^\dagger\\
&=i\begin{pmatrix}0&P_a\W h_O(u,v)P_a\\-\W h_O(v,u)&0\end{pmatrix}\\
&=\mathcal I[O]-o_{0m}\mathcal I[a^m].
\end{align*}
Consequently the explicitly constructed operator
\begin{equation}
B_O=T P_a\begin{pmatrix}\W d_O&\W q_O\\\W r_O&\W t_O\end{pmatrix}T^\dagger
\end{equation}
has the required residual.
This verifies the rate identity used above without solving for any energy eigenvector.
All participating finite polynomial and reflection operators preserve oscillator Schwartz space, which contains the energy eigenvectors and supports this compression.

\subsection{Separating relative energy shifts from intergroup mixing}

At an isolated crossing, differentiation along an integer-bias parameter and projection of the commuting relation show that its Hamiltonian derivative is diagonal in the $\pm q$ basis.
Here all operators are in the original frame, with energy projection $P_e=T^\dagger\widetilde P_eT$.
Indeed, the projected derivative of $[h,J_m]=0$ reduces to $[P_e(\partial_\alpha h)P_e,P_eJ_mP_e]=0$.
Coupling variation at fixed integer bias therefore changes the two diagonal energies without mixing their signs to first order.
Let $\Delta\alpha=\alpha-\alpha_c$, keep $\delta$ fixed, and use $d=\epsilon/\omega-m$.
At the crossing, the mixing matrix element and the two relative energy slopes are
\begin{align*}
z_{\rm mix}&=\langle\psi_+|\tau_z|\psi_-\rangle,\\
z_{\rm diff}&=\frac{\langle\tau_z\rangle_+-\langle\tau_z\rangle_-}{2},\\
a_\alpha&=\frac12\partial_\alpha(e_+-e_-).
\end{align*}
Here $e_\pm(\alpha)$ are the integer-bias energy branches at fixed $\delta$.
The subscripts distinguish intergroup mixing from the diagonal difference of the same physical bias operator $\tau_z$.
Removing a scalar energy shift, first-order degenerate perturbation theory gives the matrix in the ordered basis $(\psi_+,\psi_-)$:
\begin{equation}
h_{\rm pair}=
\begin{pmatrix}
a_\alpha\Delta\alpha+z_{\rm diff}d/2&z_{\rm mix}d/2\\
z_{\rm mix}d/2&-a_\alpha\Delta\alpha-z_{\rm diff}d/2
\end{pmatrix}
+O((|\Delta\alpha|+|d|)^2).
\end{equation}
The off-diagonal element is real in this real-doublet convention.
The factor $d/2$ comes directly from the bias term in the Hamiltonian.
Unlike a coupling change at integer bias, bias detuning mixes the two states and can shift their energies unequally.
The resulting gap is
\begin{equation}
\frac{\Delta E_{\rm gap}}{\omega}
=2\sqrt{\left(a_\alpha\Delta\alpha+\frac{z_{\rm diff}d}{2}\right)^2+
\frac{|z_{\rm mix}|^2d^2}{4}}+O((|\Delta\alpha|+|d|)^2).
\end{equation}
For $a_\alpha\ne0$, $\Delta\alpha=-z_{\rm diff}d/(2a_\alpha)$ isolates the mixing slope; at fixed $\alpha=\alpha_c$ the diagonal term remains if $z_{\rm diff}\ne0$.
The two paths thus give
\begin{equation}
\frac{\Delta E_{\rm gap}^{\rm fixed}}{\omega}
=|d|\sqrt{z_{\rm diff}^2+|z_{\rm mix}|^2}+O(d^2),
\qquad
\frac{\Delta E_{\rm gap}^{\rm comp}}{\omega}
=|d||z_{\rm mix}|+O(d^2).
\end{equation}
Compensation cancels the relative diagonal energy shift and leaves the off-diagonal coupling.
Using Eq.~\eqref{eq:gamma}, its remaining slope is $|z_{\rm mix}|=m!|\gamma|/q$, determined entirely at the undetuned crossing.
Along this local first-order path the bias remains noninteger and the surviving term $z_{\rm mix}d/2$ measures the mixing.

\subsection{Independent interference prediction and spectral test}

The main-text readout uses a specified relative phase between the two charge eigenstates.

\paragraph{Predict the response from finite partners.}

For Fig.~4, finite states are generated from Eq.~\eqref{eq:statecoeff}, displaced in the oscillator basis, and normalized as the complete partners.
The mirror amplitude is evaluated directly from $\tau_y\mathcal D(s)P_a$; $\gamma$ is its $m$th derivative at zero divided by $m!$.
The weight $q$ is computed from Eq.~\eqref{eq:square_det} and checked against the partner norm ratio.
Both quantities are determined from the integer-point states before comparison with the spectral gaps.
The common second-channel anchor has
\begin{equation}
\delta=1.3,\quad \alpha_c=0.572276157113,\quad q=3.885803614167,
\quad |\gamma|=0.357629192984,\quad \rho=0.977513538082.
\end{equation}
Its bias-response coefficients are $|z_{\rm mix}|=0.184069617765$, $z_{\rm diff}=0.340893652239$, and $a_\alpha=1.172874605876$, giving $\Delta\alpha=-0.145323997353d$.
The fixed-path slope is $0.387414643917$ and the compensated slope is $0.184069617765$.
The initial mirror orientation is chosen to make $\gamma>0$; this only interchanges the phase-preparation sign and leaves the gap prediction unchanged.
The coherent preparation and the equal incoherent mixture both have positive- and negative-group probabilities $1/2$.
Their different probe signals distinguish intergroup coherence at fixed populations.
In Fig.~4(d) the fixed-$g$ path retains the relative energy shift, while on the compensated path only mixing contributes to the linear gap.
The numerical circles retain higher-order corrections at finite detuning.

Figure~4(b) uses 301 positions in $[-2.5,2.5]$.
Panel (c) uses 31 logarithmic positions $s\in[0.008,0.12]$, away from floating-point cancellation, for $m=1,2,3$ at $\delta=1.3$.
Panel (d) samples 41 detunings $d\in[-0.025,0.025]$ and compares the spectral gaps with the separate first-order predictions, resolving higher-order deviations at finite $d$.

\paragraph{Check the predicted slope against the physical spectrum.}

An extended consistency check evaluates prediction curves at 70 values $\delta\in[0.15,1.3]$ per channel.
At 11 points per channel the original Hamiltonian is diagonalized at $d=\pm10^{-4},\pm5\times10^{-5}$ on the compensated path.
For each step the average of the two gap/absolute-detuning ratios removes the leading odd correction; Richardson extrapolation $(4s_{h/2}-s_h)/3$ then estimates the zero-detuning slope.
This extracts the spectral slope for comparison with the independently predicted amplitude.
The adjacent first-family energy indices are $m+2,m+3$, counted from zero, and are checked against the known doublet energy.

The second family tests the same prediction with a nonconstant mirror polynomial.
For $n=m=2$ and $\Delta/\omega=2.6$, the smaller positive constraint root gives $\alpha_c=0.476106812355$ and $\Phi_{2,2}(t)=1+(x+y-3)t/2$.
The integer-bias mirror overlap gives $|\gamma|=0.54536934$, and the independent charge determinant~\eqref{eq:square_det} gives $q=5.52963446$, predicting $2|\gamma|/q=0.197253307$.
The direct overlap agrees with the full profile~\eqref{eq:fullmirror} to within $10^{-12}$ on 141 equally spaced points in $[-3.5,3.5]$ and at $s=\pm2\alpha_c$.
Independently, at each $d=\pm10^{-4}$ we scan $|\alpha-\alpha_c|\le |d|$ for the minimum gap between levels 6 and 7 (counted from zero), without using the predicted compensation slope.
The mean of the two minimum gaps divided by $\omega|d|$ is $0.197253305$ at oscillator cutoffs 48, 64, and 80.

\begin{table}[ht]
\centering

\caption{Independent checks of the interference measurement and spectral predictions in main-text Fig.~4. Errors are maxima over the tested data; energies are in units of $\omega$.}

\begin{tabular}{lr}
\toprule
Check & Largest error\\
\midrule

Fig.~4(b): direct mirror curve versus polynomial shape & $5.6\times10^{-16}$\\
Finite-state eigen-equation residual & $1.1\times10^{-13}$\\
Mirror prediction $m!|\gamma|/q$ versus $|z_{\rm mix}|$ & $2.7\times10^{-15}$\\
Extended test: 33 extrapolated spectral slopes versus prediction ($N_{\rm cut}=64$) & $1.7\times10^{-10}$\\
Extended test: change of spectral slope between cutoffs 48 and 64 & $2.1\times10^{-10}$\\
Second family ($n=m=2$): minimum-gap ratio versus $2|\gamma|/q$ at $|d|=10^{-4}$ & $<1.5\times10^{-9}$\\
\bottomrule
\end{tabular}
\end{table}